\documentclass[twocolumn]{aa}  

\usepackage{amsmath}
\usepackage{atbegshi}
\usepackage{booktabs}
\usepackage{color}
\usepackage{graphicx}
\usepackage[breaklinks=true, hidelinks]{hyperref}
\usepackage{natbib}
\usepackage{pdflscape}
\usepackage{silence}
\usepackage{siunitx}
\usepackage{ulem} 
\usepackage[varg]{txfonts}
\usepackage{orcidlink}
\usepackage{placeins}
\definecolor{darkorange}{rgb}{1.0, 0.55, 0.0}
\usepackage{soul}
\begin{document} 

   \title{PDRs4All\\ XXII. Near-Infrared continuum in the Orion Bar\\}
    
\author{%
  Takashi Onaka \inst{\ref{TokyoAstro}} \orcidlink{0000-0002-8234-6747} 
  \and
  Emmanuel Dartois \inst{\ref{SaclayISM}} \orcidlink{0000-0003-1197-7143} 
  \and
 Els Peeters \inst{\ref{WOntarioPA}, \ref{WOntarioIESE}, \ref{CarlSagan}} \orcidlink{0000-0002-2541-1602} 
 \and
  Olivier Bern\'e\inst{\ref{ToulouseIRAP}} \orcidlink{0000-0002-1686-8395} 
  \and
  Emilie Habart \inst{\ref{SaclayIAS}} \orcidlink{0000-0001-9136-8043}
  \and
  Christiaan Boersma \inst{\ref{AMES}} \orcidlink{0000-0002-4836-217X} 
  \and
  Jan Cami \inst{\ref{WOntarioPA}, \ref{WOntarioIESE}, \ref{CarlSagan}} \orcidlink{0000-0002-2666-9234} 
  \and
  Asunci\'on Fuente \inst{\ref{CSIC}} \orcidlink{0000-0001-6317-6343} 
  \and
    Javier R. Goicoechea \inst{\ref{MadridFisFundamental}} \orcidlink{0000-0001-7046-4319}
    \and
    Ozan Lacinbala \inst{\ref{SaclayISM}} \orcidlink{0000-0002-5263-7922} 
    \and
    Yoko Okada \inst{\ref{Koln}} \orcidlink{0000-0002-6838-6435} 
    \and
     Alexander G.~G.~M. Tielens \inst{\ref{Leiden}, \ref{Maryland}} \orcidlink{0000-0003-0306-0028} 
     \and
     Dries Van De Putte \inst{\ref{WOntarioPA}, \ref{WOntarioIESE}} \orcidlink{0000-0002-5895-8268} 
     \and
     Francois Boulanger \inst{\ref{SorbonneLab}} \orcidlink{0000-0003-1097-6042} 
     \and
  Thomas Pino \inst{\ref{SaclayISM}} \orcidlink{0000-0002-1646-7866} 
  \and
  Yong Zhang \inst{\ref{Sunyatsen}} \orcidlink{0000-0002-1086-7922}
}

\institute{%
  Department of Astronomy, Graduate School of Science, The University of Tokyo, 7-3-1 Bunkyo-ku, Tokyo 113-0033, Japan
  \label{TokyoAstro} \and
  Institut des Sciences Mol\'eculaires d'Orsay, Universit\'e Paris-Saclay, CNRS, B\^atiment 520, 91405 Orsay Cedex, France
  \label{SaclayISM} \and
    Department of Physics \& Astronomy, The University of Western Ontario, London ON N6A 3K7, Canada
  \label{WOntarioPA}     \and 
  Institute for Earth and Space Exploration, The University of Western Ontario, London ON N6A 3K7, Canada
  \label{WOntarioIESE}          \and 
  Carl Sagan Center, SETI Institute, 339 Bernardo Avenue, Suite 200, Mountain View, CA 94043, USA
  \label{CarlSagan}         \and
    Institut de Recherche en Astrophysique et Plan\'etologie, Universit\'e Toulouse III - Paul Sabatier, CNRS, CNES, 9 Av. du colonel Roche, 31028 Toulouse Cedex 04, France
  \label{ToulouseIRAP}         \and
    Institut d'Astrophysique Spatiale, Universit\'e Paris-Saclay, CNRS,  B\^atiment 121, 91405 Orsay Cedex, France
  \label{SaclayIAS}        \and 
      NASA Ames Research Center, MS 245-6, Moffett Field, CA 94035-
   1000, USA
  \label{AMES}    \and
    Centro de Astrobiolog\'{\i}a (CSIC-INTA),Ctra de Torrej\'on a Ajaalvir, km 4, 28850, Torrej\'on de Ardoz, Spain
  \label{CSIC} \and
    Instituto de Física Fundamental (CSIC), Calle Serrano 121-123, 28006, Madrid, Spain
  \label{MadridFisFundamental} \and  
    I. Physikalisches Institut der Universit\"at zu K\"oln, Z\"ulpicher Stra{\ss}e 77, 50937 K\"oln, Germany
  \label{Koln} \and
    Leiden Observatory, Leiden University, P.O. Box 9513, 2300 RA Leiden, The Netherlands
  \label{Leiden} \and
  Astronomy Department, University of Maryland, College Park, MD 20742, USA
  \label{Maryland} \and
  Laboratoire de Physique de l'\'Ecole Normale Sup\'erieure, ENS, Universit\'e PSL, CNRS, Sorbonne Universit\'e, Universit\'e de Paris, 75005, Paris, France
  \label{SorbonneLab} \and
  School of Physics and Astronomy, Sun Yat-sen University, 2 Da Xue Road, Tangjia, Zhuhai 519000,  Guangdong Province, China
  \label{Sunyatsen} 
}

\titlerunning{NIR continuum in the Orion Bar}
\authorrunning{Onaka et al.}

   \date{received 3 March  2026 / accepted 17 July 2026}

  \abstract
   {Conspicuous excess emission is present in the near-infrared (NIR) region in various objects, including reflection nebulae, planetary nebulae, and nearby galaxies.  However, the spatial distribution and spectral shape of the excess emission remain poorly understood.} 
   {We studied the NIR continuum emission spectroscopically and obtained its spatial distribution relative to the aromatic infrared band (AIB) at 3.3\,$\mu$m in the Orion Bar prototypical photodissociation region (PDR).  We aim to characterize its spectral shape and discuss its origin.}   
   {We employed 3D spectroscopic data of the Orion Bar taken with the integrated field unit of NIRSpec on JWST from the Early Release Science program "PDRs4All." Contribution from the foreground ionized gas was estimated using the Cloudy code and subtracted.  The observed regions were divided into nine physically distinct regions and an average spectrum was derived for each region.  
   }
   {The nine regions, including the ionized gas, atomic PDR, and molecular PDR, clearly show remaining continuum in the region 1--4.5\,$\mu$m. The continuum at wavelengths longer than 2.7\,$\mu$m shows good correlations with the 3.3\,$\mu$m AIB, while the correlation of the continuum at 1.2\,$\mu$m is not significant.  We further find that the NIR continuum in the Orion Bar can be approximated by a summation of two blackbodies. The low-temperature component correlates with the AIB well, while the high-temperature component does not. The average spectra also show absorption features at 3.0 and 4.27\,$\mu$m, which are attributed to the presence in the spectra of water ice and CO$_2$ ice. }
   {We discuss possible origins of the NIR continuum, among which recurrent fluorescence from carbon clusters better explains the observed low-temperature component. The presence of ice species suggests a contribution from a deeper layer of the PDR along the line of sight producing characteristic ice absorption features.  
   }

   \keywords{Infrared: ISM
-- ISM: photon-dominated region (PDR)
-- ISM: atoms
-- ISM: lines and bands}

   \maketitle
\section{Introduction}

  The presence of conspicuous excess continuum emission in the near-infrared (NIR) region (1.5--3.7\,$\mu$m) was first reported in two reflection nebulae based on ground-based observations \citep{Sellgren1984}.  The excess NIR emission is seen in a larger sample of reflection nebulae \citep{Sellgren1996} and planetary nebulae \citep{Phipps1984, Ohsawa2016}.  It is also seen in nearby galaxies \citep{Helou2000, Lu2003} as well as in the diffuse galactic light \citep{Tsumura2013}.  AKARI observations indicate a step-increase in the NIR continuum emission across the aromatic infrared band (AIB) at the 3.3\,$\mu$m band in the reflection nebula NGC\,7023 and a molecular cloud in the Large Magellanic Cloud \citep{Boulanger2011}.  The presence of a jump between 2.5 and 3.6\,$\mu$m is also seen in JWST observations of reflection nebulae \citep{Boersma2023}. Recently, \citet{Vulcani2025} suggested that the red color excess seen in some distant galaxies may also be attributed to the excess NIR continuum emission.  

  Several models have been proposed as possible carriers of the excess NIR continuum emission, including the thermal emission from very small carbonaceous dust or carbon clusters that are stochastically heated \citep{Sellgren1984, Duley2009, Otsuka2017}, the quasi-continuum of polycyclic aromatic hydrocarbon (PAH) cations \citep{Esposito2024}, and the recurrent fluorescence of carbon clusters \citep{Lacinbala2023}.
  Since all of these suggest that the carriers are related to very small carbonaceous species in photodissociation regions (PDRs), observational studies have focused on investigating its correlation with the AIB at 3.3\,$\mu$m in the diffuse emission in our Galaxy, discussing possible origins of the emission.  
  
  \citet{An2003} studied the spatial distribution of the diffuse emission in the reflection nebula NGC 7023 using both imaging and spectroscopic data and concluded that the distribution of the NIR continuum emission is likely to be distinct from but related to the 3.3\,$\mu$m AIB distribution. On the other hand, \citet{Lu2004} analyzed {\it Spitzer} IRAC images and compared the NIR continuum emission with the mid-infrared (MIR) color of the diffuse emission of our Galaxy.  They suggested a close physical relation of the carriers of the NIR continuum emission with the AIB carriers. \citet{Flagey2006} investigated the diffuse Galactic emission based on IRAC imaging observations together with ISOCAM CVF spectroscopic data and found no clear correlation between the NIR continuum emission and the 3.3\,$\mu$m band intensity.  \citet{Haraguchi2012} studied the NIR diffuse emission in the Orion nebula based on the photometric data taken with a ground-based cryogenic telescope. They concluded that the NIR continuum emission does not show a clear correlation with the 3.3\,$\mu$m band emission after subtracting the free-free emission estimated from the extrapolation of radio data.  Recently, \citet{Peeters2024} showed that the continuum at 3\,$\mu$m has a spatial distribution similar to the 3.3\,$\mu$m AIB based on JWST observations of the Orion Bar region.  
  
  AKARI Infrared Camera (IRC) NIR spectroscopy allowed us to estimate, using hydrogen recombination lines, the contribution of the free-free emission, which amounted to about half the observed continuum flux in planetary nebulae and star-forming regions. Significant NIR continuum emission at around 3.7\,$\mu$m remained after subtracting the contribution of the ionized gas in planetary nebulae \citep{Ohsawa2016} and in Galactic star-forming regions \citep{Mori2014}.  This latter work also suggests a correlation between the ratio of the continuum to the 3.3\,$\mu$m AIB intensity and the ratio of the aliphatic-to-aromatic C-H stretching band intensities.

  While previous studies do not show correlations between the NIR continuum emission and the 3.3\,$\mu$m AIB unambiguously, they mainly employed imaging data supplemented by spectroscopic data, leaving uncertainties in their conclusions. In particular, the contribution of the free-free emission must be accurately subtracted from the data in the exact same region, where ionized gas is observed \citep[e.g.,][]{Haraguchi2012}.  Spectroscopic data in the NIR and MIR not only provide intensities of hydrogen recombination lines \citep[e.g.,][]{Peeters2024, Vandeputte2024}, from which the contribution of the free-free emission can be estimated reliably, but also allow us to study the spectral characteristics of the excess emission.

  In this study, we leveraged the 3D spectroscopic mapping data of the Orion Bar, obtained with the integral field unit (IFU) of the NIRSpec \citep{Boker2022, Jakobsen2022} on board JWST \citep{Gardner2006}, as part of the Early Release Science Program ``PDRs4All: Radiative feedback from massive stars'' \citep[ID: 1288, PIs: Bern\'e, Habart, Peeters,][]{Berne2022}, to perform a detailed and accurate study of the NIR continuum in relation to the 3.3\,$\mu$m AIB in the Orion Bar region. Thanks to JWST's high spatial resolution, we successfully studied each of the physically distinct, stratified layers of the PDR from the ionization front (IF) unambiguously \citep{Habart2024, Peeters2024}. Understanding NIR excess emission in different physical environments gives us important clues on the nature of nanometer-sized carbonaceous particles. Our observations and data reduction procedure are described in Sect.~\ref{sec:obs}. Section~\ref{sec:f-f} summarizes the estimate of the contribution from the ionized gas using the Cloudy code \citep{Ferland2017}.  Our results and analysis of the observations are presented in Sect.~\ref{sec:result} and discussed in Sect.~\ref{sec:discussion}.  Section~\ref{sec:summary} summarizes our conclusions.

\section{Observations}
\label{sec:obs}

\begin{figure}[tbh]
   \centering
   \includegraphics[width=0.7\hsize]{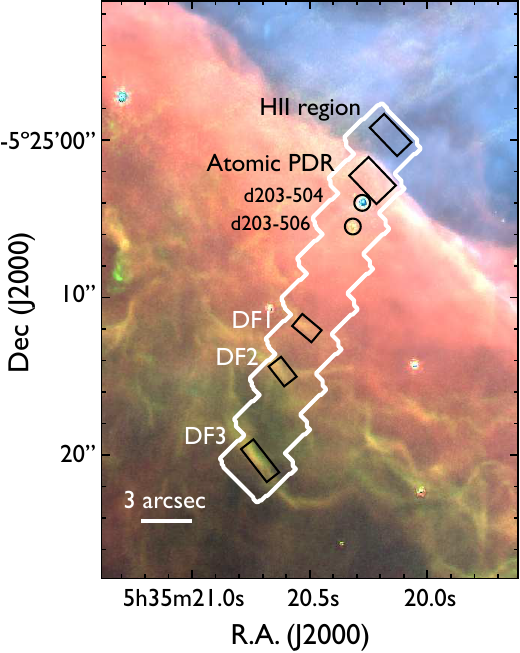}
      \caption{NIRSpec IFU footprints of the observed area of the Orion Bar (white boundaries), taken from \citet{Peeters2024}.  The regions from which the templates are extracted are indicated by the black rectangles. The two black circles show the locations of two protoplanetary disks. Red, green, and blue represent the intensities of F335M (AIB), F470N-F480M (H$_2$ emission), and F187N (Paschen $\alpha$, respectively).
              }
         \label{fig:obs}
   \end{figure}

The target of the present study, the Orion Bar, presents a prototypical PDR viewed nearly edge-on \citep[][and references therein]{Habart2024}.  The region has been extensively studied with various instruments \citep[e.g.,][]{Tielens1993, Hogerheijde1995, Goicoechea2016, Joblin2018}.  The distance to the Orion Bar is estimated as $414\pm7$\,pc \citep{Menten2007}, and the dominant illuminating star is the O7V-type star $\theta^1$ Orion C \citep{Sota2011}.  The intensity of the ultraviolet radiation field $G_\mathrm{0}$ at the IF is estimated as $G_\mathrm{0} = (2.2-7.1)\times 10^4$, where 1\,$G_\mathrm{0}$ corresponds to $1.6 \times 10^{-3}$\, erg\,cm$^{-2}$\,s$^{-1}$ between 6 and 13.6\,eV \citep{Peeters2024}.  The gas density of the atomic PDR is estimated as $(5-10) \times 10^4$\,cm$^{-3}$ \citep{Habart2024}.
NIRSpec IFU observations of a $9 \times 1$ mosaic were made to observe from the ionized region into the molecular region of a $3\arcsec \times 25\arcsec$ area with an angular resolution of $0.057\arcsec$ to $0.173\arcsec$ at a pixel size of $0.1\arcsec \times 0.1\arcsec$ (Fig.~\ref{fig:obs}). The observations employed three high spectral resolution ($R\sim 2700$) gratings (G140H, G235H, and G395H), which covered the wavelength range from 0.97 to 5.27\,$\mu$m.  We reduced the data using the JWST pipeline (version 1.10.2) with the jwst\_1084.pmap context of the Calibration Reference Data System (CRDS). Details of the NIRSpec observations, data reduction process, and NIRSpec cube production of the observed region are given in \citet{Peeters2024}.

Five templates were extracted to represent the typical spectra of the ionized region (\ion{H}{ii}\xspace), atomic PDR, as well as dissociation front 1 (DF1), 2 (DF2), and 3 (DF3) of the Orion Bar region, where DF3 is the main dissociation front (Fig.~\ref{fig:obs}). See \citet{Peeters2024} and \citet{Chown2024} for the accurate positions of the regions, where the templates were extracted. A schematic view of the observed area containing these regions is shown in \citet{Habart2024} and \citet{Peeters2024}.  In addition to these regions, there are a layer of ionized gas  \citep{ODell2020}, and closer to the observer, layers of neutral gas, known as the "Veil" \citep[e.g.,][]{Boersma2012, vanderWerf2013,Pabst2019,Pabst2020} in the foreground. Therefore, even in the molecular regions of the Orion Bar, there is a contribution from the ionized gas in the foreground, which must be corrected to study the NIR continuum emission. The Veil also causes foreground extinction\footnote{Technically speaking, this should be called attenuation. Previous studies \citep{Habart2024, Peeters2024} use the same extinction curve, and we use the term "extinction" rather than "attenuation" throughout this paper.}, which we assume to be uniform over the observed region, with $A_V = 1.5$ and $R_V=5.5$ \citep{Peeters2024}.  To account for the foreground extinction, we employed the latest extinction curve derived by \citet{Gordon2023} with $R_V=5.5$.

\section{Estimate of the ionized gas contribution}
\label{sec:f-f}
To estimate the contribution of the free-free and free-bound emission from the ionized gas, we used the Cloudy code \citep{Ferland2017}.  First, we applied the simulation results to the three templates of \ion{H}{ii}\xspace, atomic PDR, and DF3 regions to investigate their applicability.  We used the six strongest lines from the ionized gas in the Orion Bar region, \ion{H}{i}\xspace Paschen (Pa) $\alpha$, $\beta$, and $\gamma$ at 1.8756, 1.2822, and 1.0941\,$\mu$m, and Bracket (Br) $\alpha$ and $\beta$ at 4.0525, and 2.6260\,$\mu$m, respectively, and \ion{He}{i}\xspace ($^3$S--$^3$P$_0$) at 1.0834\,$\mu$m, to search for the combination of the gas density and ionizing photon flux that reproduces the observed intensities. Details of the Cloudy simulations are summarized in Appendix~\ref{app:cloudy}. Typical uncertainties are estimated at $\sim 1-3$ MJy\,sr$^{-1}$ (see Appendix~\ref{app:cloudy}).  As shown in Fig.~\ref{fig:line-fit}, the simulations reproduce the observed line intensities fairly well.

   \begin{figure*}
   \centering
   \includegraphics[width=\hsize]{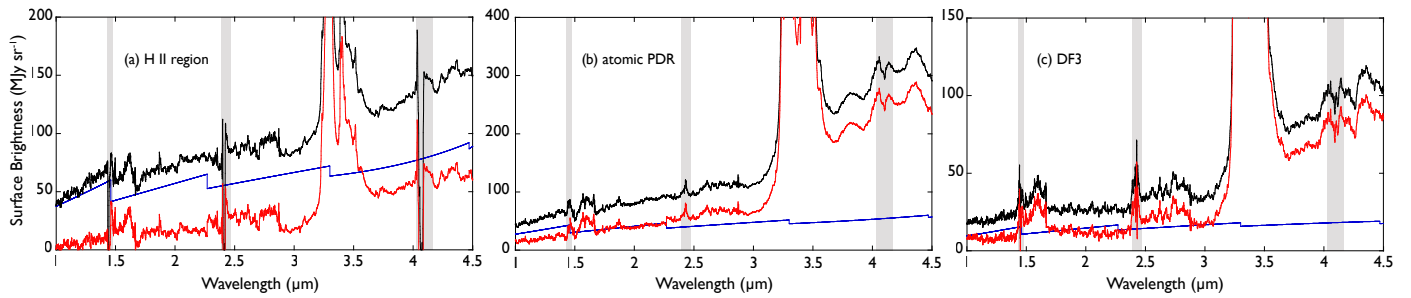}
      \caption{Observed spectra, estimated contribution from the ionized gas, and the residual continuum, shown by the black, blue and red lines, respectively, for (a) the \ion{H}{ii}\xspace region, (b) atomic PDR, and (c) DF3. The black and red spectra are smoothed by the Savitzky-Golay filter after clipping the emission lines (see text). The shaded regions indicate the spectral gaps in the observations, where the data are not reliable.   }
         \label{fig:template}
   \end{figure*}
To show the continuum clearly, emission lines were removed via sigma-clipping (3$\sigma$) using a 32-point boxcar averaging process, which is described in detail in Appendix~\ref{app:removal}. After subtracting the contribution from the ionized gas, the spectra were further smoothed by applying a 16-point Savitzky-Golay filter of a fourth-order polynomial to highlight the continuum and reduce local noise. The spectra presented in the following have these line removal and filtering processes applied. Figure~\ref{fig:template} shows the three template spectra together with the free-free and free-bound emission estimated from the parameters as determined in Appendix~\ref{app:cloudy}, as well as the remaining continuum after subtracting the contribution from the ionized gas. There is a slight mismatch at the bound-free jump at 2.3\,$\mu$m, but the amplitude of the mismatch is much smaller than the correction for the ionized gas and does not impact our subsequent analysis of the underlying continuum. The emission for wavelengths longer than 4.5\,$\mu$m is affected by the contribution of the thermal emission from sub-micrometer dust, and the following discussion focuses on the continuum from 1 to 4.5\,$\mu$m.
Figure~\ref{fig:template} indicates that the contribution of the ionized gas in the \ion{H}{ii}\xspace template is nearly 95\% at 1\,$\mu$m. Nonetheless, continuum emission remains, which becomes more evident at longer wavelengths. The contribution becomes approximately 85\% at 2\,$\mu$m and decreases to about 80\% at 3\,$\mu$m. The contribution is less in the atomic PDR and DF3 template spectra, being at a level of 50\%. The analysis of the template spectra clearly confirms excess continuum emission in both the \ion{H}{ii}\xspace region and the PDR.

We extended the same analysis to the entire area observed with NIRSpec (within the white boundaries in Fig.~\ref{fig:obs}), excluding the \ion{H}{i}\xspace Br\,$\alpha$ to consistently analyze the ionized gas contribution.
The Br\,$\alpha$ line is not available for the entire NIRSpec region due to 
the gap in spectral coverage \citep{Peeters2024}. We verified the impact of excluding Br\,$\alpha$ by comparing the results with and without it for regions where it is observed. Details are summarized in Appendix~\ref{app:cloudy} and Table~\ref{table:compBra}. As shown in Table~\ref{table:compBra},  excluding Br\,$\alpha$ introduces an average difference in the continuum of $\sim 0.3-0.5$\,MJy\,sr$^{-1}$.  The average difference relative to the results without Br\,$\alpha$ is only a few percent. The exclusion affects the estimate of the continuum level more at shorter wavelengths, where the fractional contribution of free-free and free-bound emission is greater. These values are a factor of 4--6 smaller than the uncertainties in the line fit described above. The pixel-level difference shows large scatter, but the general trend is not affected. Therefore, the exclusion does not significantly impact the following correlation analysis (see Sect.~\ref{subsec:map} and Table~\ref{table:compBra}). We applied the same analysis to the entire region covered by NIRSpec, where all the five line intensities used in the fit have a signal-to-noise ratio (S/N) larger than three, and subtracted the estimated free-free and free-bound emission from the observed spectrum. Spatial pixels (spaxels), where the S/N is less than three, are excluded from the following analysis.
 
\section{Results}
\label{sec:result}

\subsection{Average spectrum across the Orion Bar}
\label{subsec:spectrum}

   \begin{figure*}
   \centering
   \includegraphics[width=0.9\hsize]{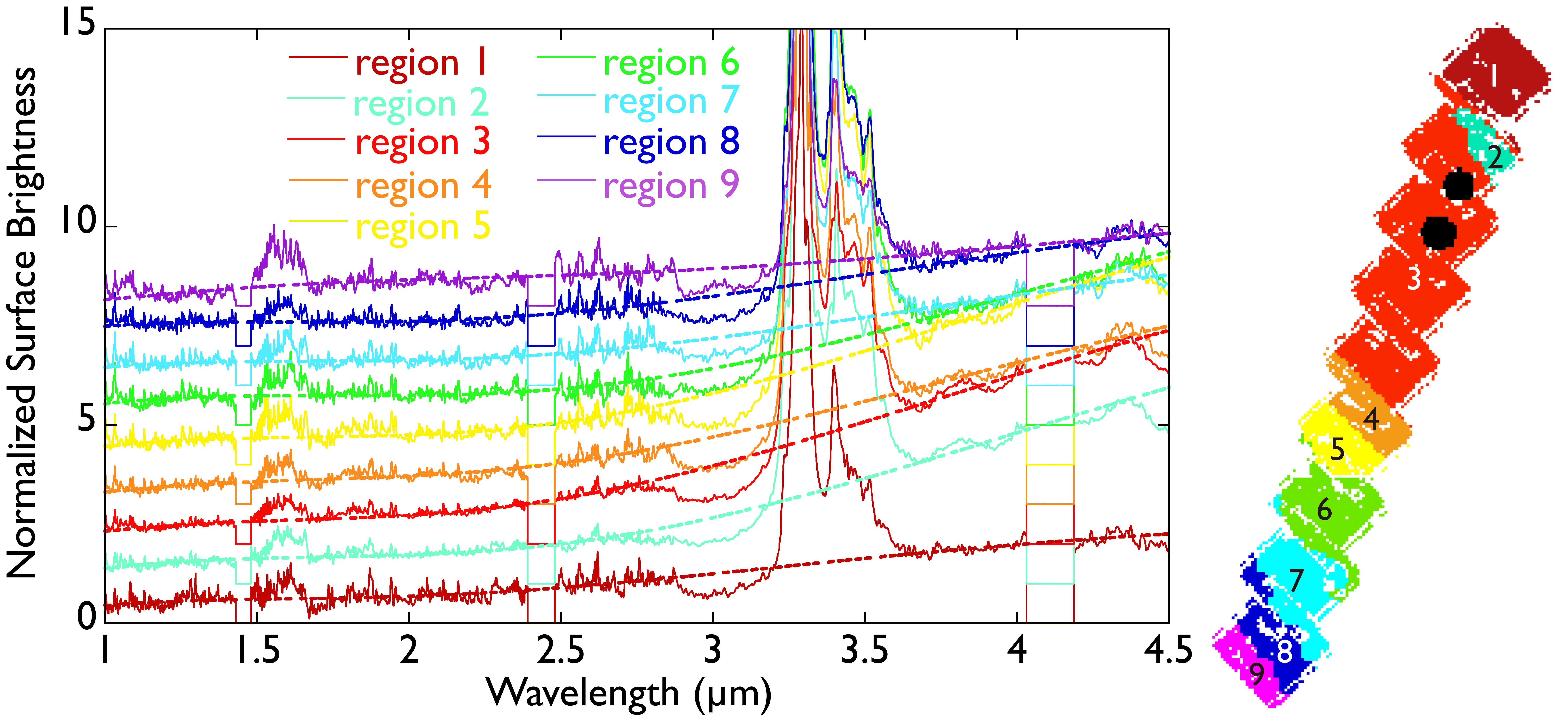}
      \caption{Left: Extinction-corrected average spectra of the nine regions (right panel) after subtracting the contribution from the ionized gas. Emission lines were removed via sigma-clipping (3$\sigma$), and the spectra were smoothed by the Savitzky-Golay filtering  (see text for details). The solid lines show the spectrum in each region normalized at the average surface brightness between 2.6 and 2.7\,$\mu$m.  They are shifted by unity to show them separately. The spectra at around the 3.3\,$\mu$m AIB are clipped to show the continuum clearly.  The spectral regions that show zero intensity correspond to the gaps in the spectra due to the grating setting.  Right: Allocation of the regions defined in
      \citet{Khan2025}.  Regions 1, 2, ..., and 9 correspond to the ionized region (brown), teal region (teal) located just below the IF, atomic PDR (red), DF1 (orange), region between DF1 and DF2 (yellow), DF2 (green), region between DF2 and DF3 (light blue), DF3 (blue), and region beyond DF3 (purple), respectively. The large black circles indicate the regions around two protoplanetary disks, d203-504 and d203-506, excluded from the present analysis. The white spots indicate the spatial pixels. These were excluded from our analysis, because either the S/Ns of the line intensities are below three or the data contain a spectral point with deviations exceeding 5$\sigma$ from the average spectrum.} The dashed lines show the spectra of the two-blackbody fit (see Sect.~\ref{subsec:two-component}). 
         \label{fig:average}
   \end{figure*}

After subtracting the contribution from the ionized gas from each spectrum, we normalized the residual continuum by the average intensity between 2.6 and 2.7\,$\mu$m.  This wavelength range was chosen to exclude intense
features. Following \citet{Khan2025}, we subsequently divided the entire NIRSpec area into nine physically distinct regions described below and calculated an average spectrum for each region. We excluded spectra with deviations exceeding 5\,$\sigma$ at any spectral point to eliminate noisy spectra. This procedure removed spectra severely affected by $1/f$ noise efficiently \citep{Birkmann2022}.  Spectra at these positions are also excluded from the following analysis.  

The regions are highlighted in different colors in the right panel of Fig.~\ref{fig:average}.  Region 1 (brown) is the ionized region, where the emission from the background PDR is also observed. 
Region 2 (teal) is located just below the IF and defined as the teal region in \citet{Khan2025}. Here, the AIBs in the 10--14\,$\mu$m range show a different behavior compared to those in the atomic PDR (region 3, red), and intense \ion{O}{i}\xspace fluorescent lines are observed \citep{Peeters2024}. The two black circles in region 3 indicate where the continuum emission is strongly affected by two protoplanetary disks, d203-504 and d203-506 \citep{Berne2024, Schroetter2025}, respectively, whose data are excluded from the present analysis. Region 4 (orange) corresponds to DF1, and region 5 (yellow) is defined as the 
region between DF1 and DF2. Region 6 (green) is DF2.  The small region west of DF2 defined in \citet{Khan2025} has only a few pixels in the NIRSpec area and is excluded from the present analysis. Regions 7 (light blue), 8 (blue), and 9 (purple) correspond to the region between DF2 and DF3, DF3, and the region beyond DF3, respectively. The white spots indicate the data excluded from the following analysis because either the S/Ns of the line intensities are below three or the data contain a spectral point with deviations exceeding 5\,$\sigma$ from the average spectra.

The left panel of Fig.~\ref{fig:average} displays the average spectrum of each region, for which the foreground extinction is corrected to show their intrinsic spectra.  
Similarly to the template spectra (Fig.~\ref{fig:template}), emission lines were removed by 3$\sigma$ clipping of a 32-point boxcar average, and the spectra were smoothed by applying a 16-point Savitzky-Golay filter of a fourth-order polynomial. The spectral regions with zero intensity correspond to
gaps in the spectrometer coverage.   
The extinction effect is small \citep[$A_\mathrm{V}$ = 1.5,][]{Peeters2024}, and the free-free and free-bound emission was derived from the emission lines that cover a restricted spectral range. Therefore, the extinction correction has no effect on the results of the ionized gas contribution subtraction. The correction affects the slope only slightly and does not affect the following analysis.  The spectra are shifted by a unit increment for the sake of clarity. The average spectra of all nine regions clearly show residual continuum emission in the 1--4.5\,$\mu$m range.  They show similar
spectral shapes at wavelengths shorter than 3\,$\mu$m, while the spectra at wavelengths longer than the 3.3\,$\mu$m AIB show differences in intensity.
Some spectra (e.g., regions 1, 3, and 4) show a shallow dip at 3\,$\mu$m.  This feature is discussed in Sects.~\ref{subsec:absorption}.
With closer inspection, each spectrum shows a weak bump around 1.6\,$\mu$m.  The feature is also seen in the template spectra (Fig.~\ref{fig:template}).  This spectral range falls on the NRS2 detector on the long-wavelength side of the spectral gap and is known to exhibit higher noise than the short-wavelength side, which falls on the NRS1 detector \citep{Peeters2024}. Therefore, some noise may persist in the spectra. We reprocessed this spectral segment with the \texttt{clean\_flicker\_noise} tool included in newer versions of the pipeline (1.20.1).
This significantly suppresses the 1.6\,$\mu$m excess, while the spectrum remains mostly unchanged at the other wavelengths. The impact of the remaining noise on this feature requires further investigation and is not discussed here.

\subsection{Continuum emission map}
\label{subsec:map}

\begin{figure*}[tbh]
   \centering
   \includegraphics[width=0.68\hsize]{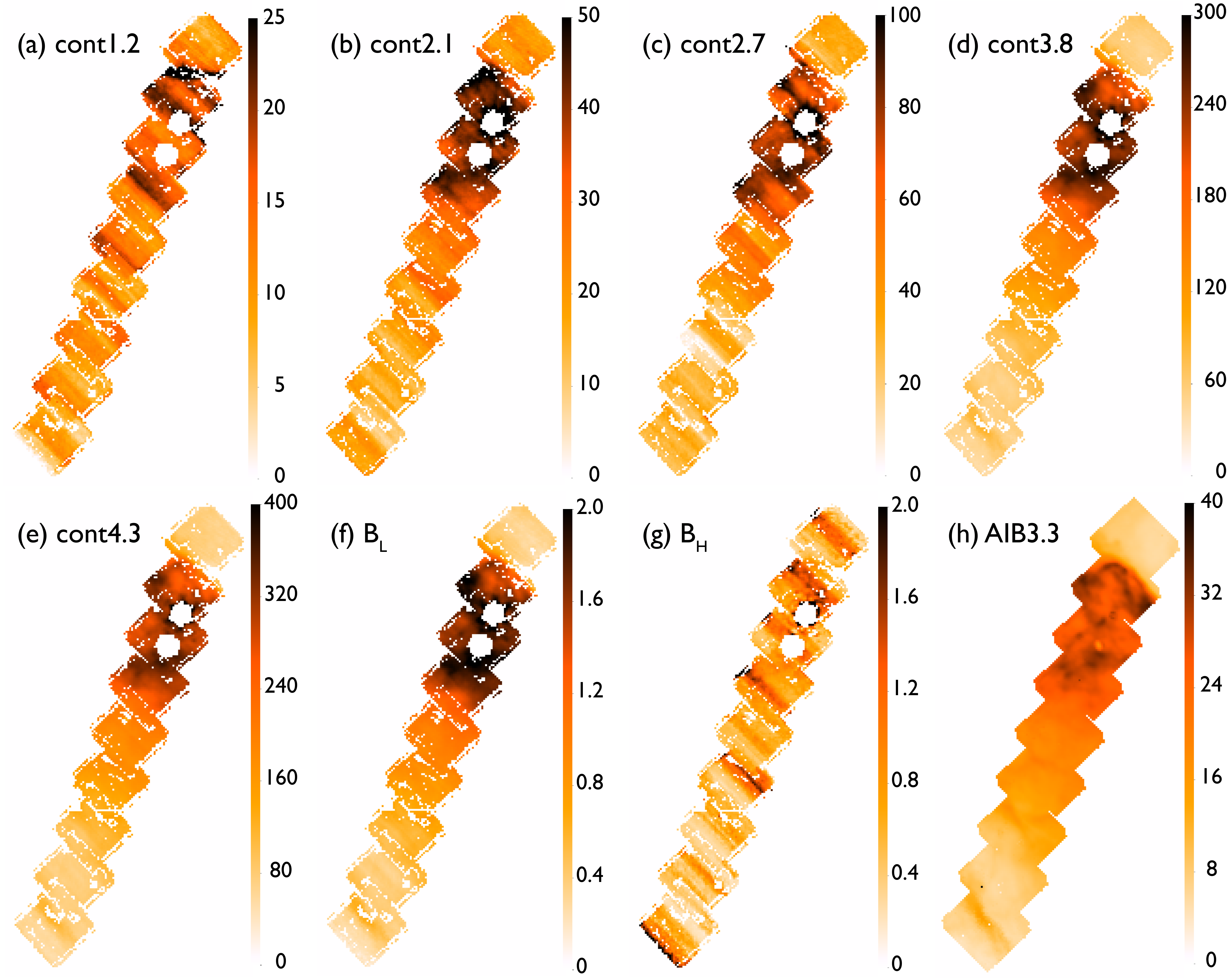}
      \caption{Mean surface brightness maps of the continuum emission at 1.2\,$\mu$m (a), 2.1\,$\mu$m (b), 2.7\,$\mu$m (c), 3.8\,$\mu$m (d),  4.3\,$\mu$m (e), the low-temperature $B_\mathrm{L}$ (f) and high-temperature $B_\mathrm{H}$ (g) components, and the 3.3\,$\mu$m AIB intensity (h).  The 3.3\,$\mu$m AIB intensity is taken from \citet{Peeters2024}.  The color bars are expressed in MJy\,sr$^{-1}$ for (a), (b), (c), (d), and (e) and in units of $\times 10^{-4}$, $\times 10^{-5}$, and $\times 10^{-6}$\,W\,m$^{-2}$\,sr$^{-1}$ for (f), (g), and (h), respectively.}  
         \label{fig:map}
   \end{figure*}

To study the spatial distribution and the correlation of the NIR continuum emission with the 3.3\,$\mu$m AIB, we calculated the mean surface brightnesses between 1.1--1.3, 2.0--2.2, 2.6--2.8, 3.7--3.9, and 4.2--4.4\,$\mu$m at each spatial position, denoted by Cont1.2, 2.1, 2.7, 3.8, and 4.3 in the following, respectively\footnote{There is a small absorption feature at 4.27\,$\mu$m (Sect.\ref{subsec:absorption}), but its contribution to Cont4.3 is insignificant.}.
Figure~\ref{fig:map} shows the mean surface brightness maps at five wavelengths and the intensity of the 3.3\,$\mu$m AIB taken from \citet{Peeters2024}.  For the data shown in Fig.~\ref{fig:map}, the foreground extinction is not corrected. The surface brightness maps at wavelengths shorter than 2\,$\mu$m are relatively noisy, but all the figures indicate that they are stronger at the atomic PDR and gradually weaken toward the molecular region, suggesting qualitatively similar spatial distributions to the 3.3\,$\mu$m AIB. The maps at 1.2, 2.1, and 2.7\,$\mu$m show noise patterns aligned with one of the IFU axes and are considered carefully in the following analysis.

Figure~\ref{fig:correlation} shows correlation plots of the mean surface brightnesses at 1.2, 2.1, 2.7, 3.8, and 4.3\,$\mu$m with the 3.3\,$\mu$m AIB intensity. The Spearman correlation coefficient for each region indicated in Fig.~\ref{fig:template} is summarized in Table~\ref{table:correlation}.  As seen in Fig.~\ref{fig:correlation}, the correlation coefficients with the 3.3\,$\mu$m AIB are large for Cont3.8 and Cont4.3 in all regions. The slightly lower value of Cont4.3 in region 9 may be attributed to its fainter continuum emission. The correlation gradually becomes weaker toward shorter wavelengths. Cont2.7 correlates relatively well with the 3.3\,$\mu$m AIB except in regions 1 and 7.  Cont2.1 correlates relatively well in regions 2, 3, 6 and 7, but in other regions, the correlation coefficients are all lower than 0.4.  In particular, they are negative in regions 4 and 8.
Cont1.2 shows correlation coefficients greater than 0.4 only in regions 3 and 9.  
If the correlation in each region for all the continuum data is considered, regions 3 and 6 show relatively strong correlations across all continuum emission.  
Region 9 shows a fair correlation, except for Cont2.1. Region 1 does not correlate well, except for Cont 3.8 and Cont4.3. The Spearman coefficients shown in Table~\ref{table:correlation} differ slightly from the derived values, including those for \ion{H}{i}\xspace Br\,$\alpha$ (Table~\ref{table:compBra}). However, this variation results from the difference in the sample area since Br\,$\alpha$ is not observed in the entire NIRSpec region (Appendix~\ref{app:cloudy}). A comparison of the same region reveals that the difference is small (Table~\ref{table:compBra}), and the general trend in the correlation does not change upon exclusion of Br\,$\alpha$.

   \begin{figure*}
   \centering
   \includegraphics[width=0.9\hsize]{correlation4.pdf}
      \caption{Correlation plots against the 3.3\,$\mu$m AIB intensity adopted from \citet{Peeters2024}. (a)  Continuum at 1.2\,$\mu$m. (b) Continuum at 2.1\,$\mu$m. (c) Continuum at 2.7\,$\mu$m. (d) Continuum at 3.8\,$\mu$m. (e) Continuum at 4.3\,$\mu$m. (f) Integrated intensity of the low-temperature component $B_\mathrm{L}$. (g) Integrated intensity of the high-temperature component $B_\mathrm{H}$ (Eqs.~(\ref{eq2}) and (\ref{eq3})). The colored dots correspond to each region shown in the right of Fig.~\ref{fig:average}.}  
         \label{fig:correlation}
   \end{figure*}
%

\begin{table*}
\caption{Spearman correlation coefficient with the 3.3\,$\mu$m AIB intensity.}             
\label{table:correlation}      
\centering                          
\begin{tabular}{c c c c c c c c c}        
\hline\hline                 
Region \#& Region & Cont1.2 & Cont2.1 & Cont2.7 & Cont3.8 & Cont 4.3 & $B_\mathrm{L}$ & $B_\mathrm{H}$ \\    
\hline                        
   1 & ionized region & 0.29 & 0.10 & 0.22 & 0.79 & 0.75  & 0.72 & -0.25\\ 
   2 & teal region & -0.03 & 0.46 & 0.45 & 0.76 & 0.72  & 0.77 & 0.34 \\
   3 & atomic PDR &  0.45 & 0.78 & 0.62 & 0.86 & 0.88  & 0.87 & 0.43 \\
   4 & DF1 & 0.18 & -0.38 & 0.55 & 0.88 & 0.85  & 0.75 & -0.34 \\
   5 & DF1 -- DF2 &  0.31 &  0.10 & 0.85 & 0.96 & 0.96 & 0.93 & -0.03\\ 
   6 & DF2 &  0.27 & 0.66 &  0.50 &  0.94 & 0.94  & 0.91 & 0.60\\
   7 & DF2 -- DF3 &  -0.09 & 0.42 & 0.10 & 0.70 & 0.64  & 0.47 & 0.02\\
   8 & DF3 &  0.03 & -0.15 & 0.47 & 0.94 & 0.92 &  0.90 & -0.12\\
   9  & beyond DF3 &  0.64 & 0.32 & 0.62 & 0.90 & 0.60  & 0.86 & -0.59\\ 
   \hline
\end{tabular}
\end{table*}

\subsection{Two-blackbody fit}
\label{subsec:two-component}
To further investigate the nature of the NIR continuum emission, we fit the continuum emission at the five wavelengths, for which we estimated the mean surface brightness described in Sect.~\ref{subsec:map}, using blackbodies after extinction-correction. We find that a single blackbody fit does not work well, so we performed the fit with a summation of two blackbodies given by
\begin{equation}
    \label{eq1}
    f_\nu(\lambda_i) = a_\mathrm{L}\,B_\nu(\lambda_i, T_\mathrm{L}) + a_\mathrm{H}\,B_\nu(\lambda_i, T_\mathrm{H}) \,, 
\end{equation}

\noindent
where $B_\nu(\lambda, T)$ is the Planck function of a temperature $T$ at a wavelength $\lambda$,  $T_\mathrm{L}$ and $T_\mathrm{H}$ denote the temperatures of the low- and high-temperature components, and $a_\mathrm{L}$ and $a_\mathrm{H}$ are the corresponding coefficients. The suffix $i$ ($=1, 2, ..., 5$) of the wavelength corresponds to 1.2, 2.1, 2.7, 3.8, and 4.3\,$\mu$m.  We searched for the best fit in the temperature range 100 -- 4000\,K for both $T_\mathrm{L}$ and $T_\mathrm{H}$.

The two-blackbody fit for the average spectrum of each region is shown in Fig.~\ref{fig:average} as dashed lines. The fit-derived temperatures, $T_\mathrm{L}$ and $T_\mathrm{H}$, are listed in Table~\ref{table:2BB}.  The general spectral shape of the continuum emission is reproduced sufficiently well by a summation of two blackbodies from 1 to 4.5\,$\mu$m.  We verified whether the two-blackbody fit is consistent with the longer-wavelength MRS spectra. For this, we used the template spectra, extracted from the same area \citep{Chown2024}, as the pixel scale of NIRSpec and MIRI/MRS is not the same. Note that the continuum was subtracted from the spectra presented in \citet{Chown2024}. The results are shown in Fig.~\ref{fig:N+M}.  The spectra rise sharply after 5\,$\mu$m, and the two-blackbody fit remains well below. Therefore, the two-blackbody fit does not conflict with the MRS spectra. It should also be emphasized that the two-blackbody fit is a phenomenological decomposition used to characterize the spectral shape of the NIR continuum. It does not ascribe any physical meaning to the emission. The fitted spectra reveal more evident absorption at 3\,$\mu$m. The results also suggest weak absorption at 4.27\,$\mu$m, except for regions 1 and 9. The contributions of the two components in the nine average spectra are shown in Fig.~\ref{fig:BBsp}, which indicates that the crossing wavelength of the two components is at around 2.5--2.7\,$\mu$m except for region 9, where the low-temperature component is relatively weak and crossing occurs at around 3.5\,$\mu$m.

%
\begin{table}
\caption{Temperatures of the two-blackbody fit.}             
\label{table:2BB}      
\centering                          
\begin{tabular}{c c c c }        
\hline\hline                 
Region \#& Region & $T_\mathrm{L}$ (K) & $T_\mathrm{H}$ (K) \\    
\hline                        
   1 & ionized region & 777 & 3315  \\ 
   2 & teal region &  617 & 2868  \\
   3 & atomic PDR &  657 & 2743  \\
   4 & DF1 & 650 & 2660  \\
   5 & DF1 -- DF2 &  632 & 3077  \\ 
   6 & DF2 &  595 & 3234  \\
   7 & DF2 -- DF3 &  672 & 3298  \\
   8 & DF3  & 700 & 3581 \\
   9  & beyond DF3 & 585 & 2106  \\ 
\hline                                   
\end{tabular}
\end{table}

 As the two-blackbody model fit the observed average spectra well, we applied it to the extinction-corrected continuum data at the five wavelengths for the entire spectroscopic region. The four fitted parameters ($T_\mathrm{L}$, $T_\mathrm{H}$, $a_\mathrm{L}$, and $a_\mathrm{H}$) were determined at each pixel.  The residuals are sufficiently small for all the pixel data.  Subsequently, the two blackbodies were integrated over the spectral range used for the continuum levels at five wavelengths (1.1 -- 4.4\,$\mu$m) to obtain the intensities of the two components $B_\mathrm{L}$ and $B_\mathrm{H}$ at each spatial pixel as

\begin{equation}
    \label{eq2}
    B_\mathrm{L} = a_\mathrm{L}\,\int_{\lambda_1} ^{\lambda_2}B_\nu(\lambda, T_\mathrm{L})\,d\lambda 
\end{equation}

\noindent and

\begin{equation}
   \label{eq3}
    B_\mathrm{H}  =  a_\mathrm{H},\int_{\lambda_1} ^{\lambda_2}B_\nu(\lambda, T_\mathrm{H})\,d\lambda ,
\end{equation}

\noindent where we set $\lambda_1 = 1.1$\,$\mu$m and $\lambda_2 = 4.4$\,$\mu$m. Figure~\ref{fig:map} shows the spatial distribution of $B_\mathrm{L}$ (Fig.~\ref{fig:map}f) and $B_\mathrm{H}$ (Fig.~\ref{fig:map}g).  Their correlations with the 3.3\,$\mu$m AIB intensity are also plotted in Figs.~\ref{fig:correlation}f and g. The Spearman correlation coefficients are given in Table~\ref{table:correlation}.  As expected from the strong correlations between the continuum emission for wavelengths longer than 2.7\,$\mu$m and the 3.3\,$\mu$m AIB, the intensity of the low-temperature component $B_\mathrm{L}$ correlates with the 3.3\,$\mu$m AIB intensity well, while the high-temperature component $B_\mathrm{H}$ correlates very weakly. Note that the map of $B_\mathrm{H}$ also shows a noise pattern similar to those seen in shorter wavelength maps (Fig.~\ref{fig:map}). Some of the structures seen in the correlation plot (Fig.~\ref{fig:correlation}g) arise from the noise pattern.  Figure~\ref{fig:Tdist} shows the distributions of $T_\mathrm{L}$ and $T_\mathrm{H}$.  $T_\mathrm{L}$ ranges between 200--1000\,K, with a mean of $\sim 700$\,K.  A large scatter in $T_\mathrm{L}$ is seen in the \ion{H}{ii}\xspace region, which may be attributed to the weaker continuum emission at longer wavelengths (see Fig.~\ref{fig:template}).  There seems to be no clear trend in $T_\mathrm{L}$ with distance from the IF and $T_\mathrm{H}$ does not show a clear spatial variation, either. However, a nonnegligible fraction of the fit results in $T_\mathrm{H}= 4000$\,K, which is the search limit. Also, a slight difference higher than 2000\,K does not significantly change our results. Therefore, its value is not well constrained. This is partly due to the faintness of the surface brightness and to the larger noise at shorter wavelengths. We applied a power-law fit for the high-temperature component, but the fit was worse than the blackbody fit. The present dataset does not constrain the shape of the high-temperature component very well, except that it exhibits a decreasing trend with wavelength.

\section{Discussion}
\label{sec:discussion}
The present study clearly shows a correlation of the NIR continuum emission at wavelengths longer than 2.7 \,$\mu$m with the 3.3\,$\mu$m AIB in the Orion Bar based on the spectroscopic data. It also shows that the continuum at $1 - 4.5$\,$\mu$m can be decomposed into two blackbodies. The low-temperature component correlates with the AIB well, while the high-temperature component does not. The different correlations suggest that the two components have different origins.
The low-temperature component is characterized by a spectrum that begins to decline across the 3.3\,$\mu$m AIB, potentially producing a jump-like structure, as previously indicated \citep{Boulanger2011, Boersma2023}.  The shape of the high-temperature component is not well constrained. In the following subsection, we discuss the possible origin of the NIR continuum and, in particular, the low-temperature component.

\subsection{Origin of the NIR continuum}\label{subsec:origin}
Emission from highly vibrationally excited molecules can produce a quasi-continuum that spans the entire MIR range \citep{Allamandola1989, Allamandola2021}.
\citet{Esposito2024} suggest that PAH cations emit a large number of lines between $3.5 - 5$\,$\mu$m, which can produce the quasi-continuum.
They only provide a spectrum of the quasi-continuum in the abstract, which appears relatively flat for 3.5--5\,$\mu$m with some structures. It is not possible to directly compare the predicted spectrum with the present results.  The present observations suggest that the low-temperature component continues down to 2.1\,$\mu$m.  If PAH cations also produce emission at wavelengths shorter than 3.5\,$\mu$m and the predicted spectrum can be approximated by a blackbody, it could be a vital candidate for its origin. The degree of ionization is expected to decrease with distance from the IF \citep{Maragkoudakis2026, Khan2026}; however, the correlation plot suggests no such trend with distance in $B_\mathrm{L}$ (Fig.~\ref{fig:correlation}g).

\begin{figure}
   \centering
   \includegraphics[width=0.9\hsize]{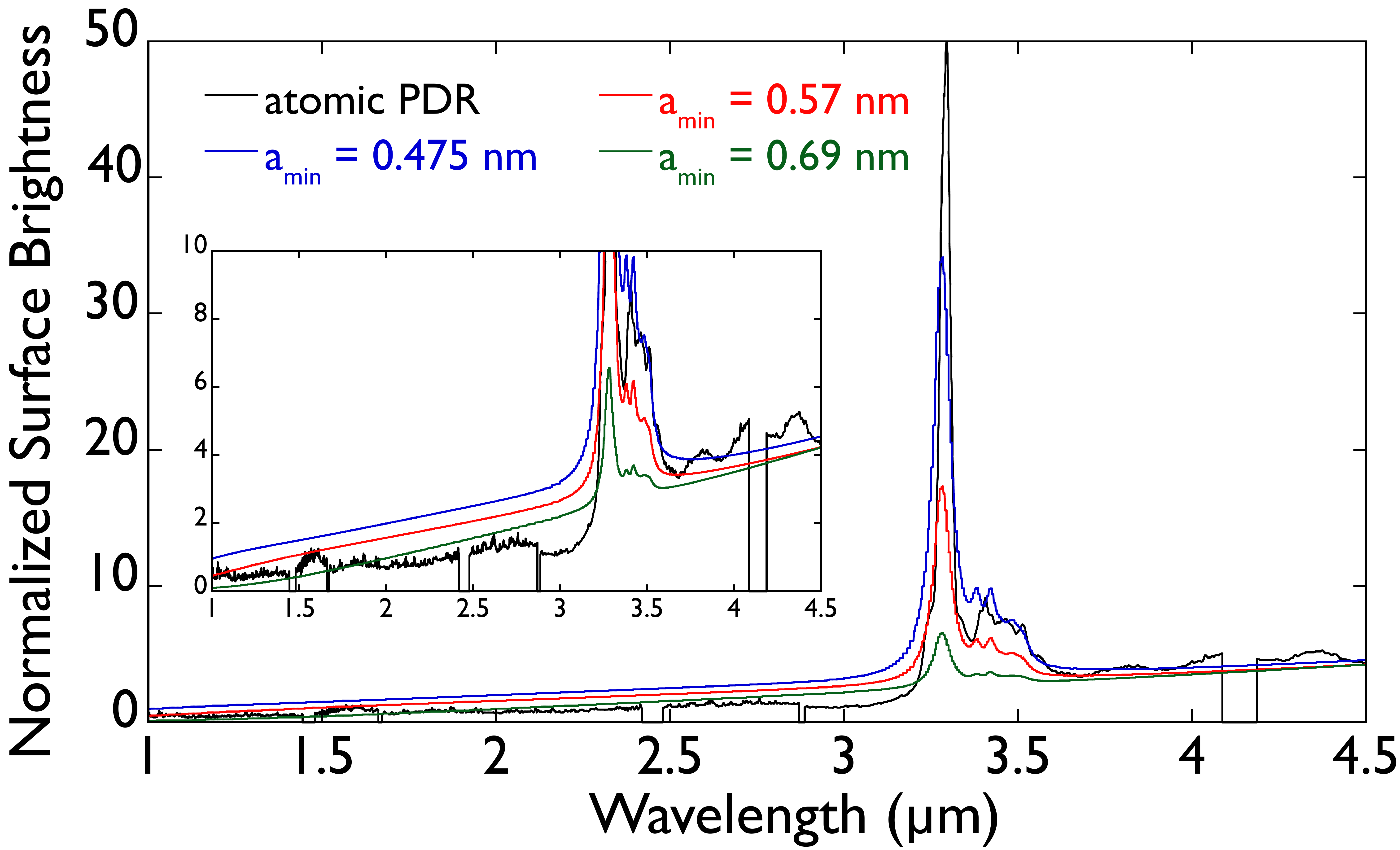}
      \caption{Comparison of the observed spectrum of the atomic PDR (black line) with those of the THEMIS model \citep{Elyajouri2024}. The model spectrum assumes that $E_\mathrm{g} = 0.03$\,eV.  The blue, red, and green lines show the spectra with $a_\mathrm{min} = 0.475$, 0.57, and 0.69\,nm, respectively. The vertical scale is expanded in the inserted figure to show the difference in the continuum clearly.}  
         \label{fig:DustEM}
   \end{figure}

Nanometer-sized carbonaceous dust grains also produce strong continuum emission in the NIR.  Here, we used The Heterogeneous dust Evolution Model for Interstellar Solids (THEMIS) \citep{Jones2017} to investigate if it could explain the observed spectrum.  
The THEMIS model includes the following two parameters: the energy gap $E_\mathrm{g}$, which determines the aliphatic-to-aromatic feature ratio, and the minimum size of nanometer-sized dust, $a_\mathrm{min}$, which affects the 3.3\,$\mu$m AIB intensity. From the template spectra, \citet{Elyajouri2024} derive $E_\mathrm{g}$ = 0.03\,eV and $a_\mathrm{min} = 0.475$\,nm for the atomic PDR in the Orion Bar. Figure~\ref{fig:DustEM} shows the spectrum of region 3 (atomic PDR; black line) and the THEMIS model spectrum from \citet{Elyajouri2024} (blue line).  
As seen in the figure, if the adopted model spectrum is scaled to the intensity of the AIBs at 3.3 and 3.4--3.5\,$\mu$m region (blue line), the continuum at wavelengths longer than 3.5\,$\mu$m matches our observations well.  However, the model significantly overestimates the continuum emission at wavelengths shorter than 3\,$\mu$m as shown in the inserted figure. Increasing $a_\mathrm{min}$ reduces the discrepancy in the continuum (red and green lines). However, in these cases, the model cannot reproduce the 3.3\,$\mu$m AIB emission sufficiently. 
The nanometer-sized dust in the THEMIS model is responsible for both the band and continuum emission, and the dust continuum emissivity does not have a sharp decline below the 3.3\,$\mu$m band. Therefore, it is difficult to reproduce a spectrum with strong 3.3\,$\mu$m band emission and continuum with a decline in the shorter-wavelength side simultaneously with the same temperature. The same problem occurs in other dust models that assume the 3.3\,$\mu$m band and continuum from the same nanometer-sized carbonaceous dust \citep[e.g.,][]{Draine2001}. However, we note that the present dust models are not optimized to reproduce the NIR continuum emission by design. The present study simply indicates that, while the 3.3\,$\mu$m AIB and the long wavelength continuum ($> 3.3$\,$\mu$m) may originate in the same component, 
the same cannot be said for the shorter wavelength continuum. Revision of the continuum emissivity of the dust model needs to be considered.

The NIR continuum emission can also arise from recurrent fluorescence of molecules \citep{Leger1988}, wherein a carrier molecule is excited to an electronic energy level by absorbing an optical to ultraviolet photon. It then undergoes internal conversion to a vibrationally excited, electronic ground state, followed by inverse internal conversion and fluorescence emission to return to the ground state. Recent experiments detect recurrent fluorescence from various PAHs \citep[e.g.,][and references therein]{Stockett2023, Bernar2023, Navarret2023, Lee2023}.  Recurrent fluorescence from small carbon clusters has also been reported in recent experiments \citep{Bull2025, Pedersen2025}.  \citet{Lacinbala2023PhRvA} calculated the spectra of recurrent fluorescence from isomers of the small carbon clusters C$_{42}$ and C$_{60}$ and show that they produce modified blackbody-like emission, which can be approximated by $\nu^2 B_\nu(T)$. These carbon clusters have low-lying electronic states, for which the vibrational energy acts as a heat bath, and the large number of isomers results in a broad and smooth emission spectrum. The additional $\nu^2$ factor originates from the electronic density of states of the carbon cluster population and the electric dipole transition moment \citep{Lacinbala2023PhRvA}.  

To investigate whether the low-temperature component can be explained by recurrent fluorescence from carbon clusters, we replaced
the low-temperature component with a modified blackbody and fit the average spectra according to the following equation:

\begin{equation}
    \label{eq4}
    f^\mathrm{M}_\nu(\lambda_i) = a^\mathrm{M}_\mathrm{L}\,\lambda^{-2}\,B_\nu(\lambda_i, T^\mathrm{M}_\mathrm{L}) + a^\mathrm{M}_\mathrm{H}\,B_\nu(\lambda_i, T^\mathrm{M}_\mathrm{H}) \,, 
\end{equation}

\noindent
where $a^\mathrm{M}_\mathrm{L}$ and $a^\mathrm{M}_\mathrm{H}$ are the fit coefficients and $T^\mathrm{M}_\mathrm{L}$ and
$T^\mathrm{M}_\mathrm{H}$ are the fitted temperatures for the low- and high-temperature components, respectively.  The fit curves are also plotted in Fig.~\ref{fig:BBsp} with the blue lines. The fitted curve is very similar to the two-blackbody fit (Eq.~(\ref{eq1})), and the difference between the two fits is insignificant because the exponential decrease dominates at the wavelength range in which the fit was performed. The values of $T^\mathrm{M}_\mathrm{L}$ are lower than $T_\mathrm{L}$, with the mean value being $\sim 500$\,K.  \citet{Lacinbala2023} estimate that recurrent fluorescence from C$_{60}$ isomers in NGC 7023, whose central star has a temperature of 20000\,K, can be approximated by a modified blackbody of 747\,K. The $T^\mathrm{M}_\mathrm{L}$ values obtained in the present study are lower than 700\,K, suggesting that larger carbon clusters, such as C$_{72}$, C$_{82}$, or C$_{90}$, could produce the emission of the low-temperature component observed in the Orion Bar. Larger clusters have more energy levels, resulting in lower "average" excitation temperatures \citep{Lacinbala2023PhRvA}.  \citet{Otsuka2017} suggest that the NIR continuum emission in a planetary nebula in the Small Magellanic Cloud can be attributed to stochastically heated small carbon molecules of 39 carbon atoms. They also note that such carbon molecules do not have a sufficient number of vibrational modes to produce a blackbody-like emission. Large carbon clusters have a large number of isomers. If these isomers are present without significant selection effects due to their formation environment or local physical conditions favoring particular forms, the emission from recurrent fluorescence can become blackbody-like, as observed. The present study shows strong correlation between the low-temperature component and the 3.3\,$\mu$m AIB intensity. The 3.3\,$\mu$m band is thought to arise from aromatic C-H stretching modes. Carbon clusters are not expected to have appreciable amounts of C-H bonds. If hydrogen attachment produces C-H bonds, they may also exhibit a vibrational mode at 3.3\,$\mu$m, which would compete with recurrent fluorescence.  The detection of trace amounts of small hydrocarbons in the atomic region suggests that carbon clusters may form through photoprocessing of PAHs  \citep{Goicoechea2025}.  
The investigation carried out in \citet{Lacinbala2023} can be transposed to any PAH or PAH-like population. Blackbody-like infrared emission of recurrent fluorescence can be produced if a large diversity of PAH species participate and their electronically excited energy levels are significantly populated. Recurrent fluorescence may be a significant contributor to the NIR continuum emission observed in the Orion Bar.

The fitted temperature of the low-temperature component, $T_\mathrm{L}$, does not show clear spatial variation.  The intensity of the incident radiation field changes with distance from the exciting source, but the spectral energy distribution (SED) of the incident radiation on the surface of the PDR should not vary significantly over the region. The nature of the quasi-continuum originating from PAH cations requires further study but is said to arise from IR fluorescence, whose spectrum depends on the energy of the exciting photon rather than the intensity of the incident radiation \citep{Allamandola1989}.  The emission from the nanometer-sized dust grains can be described by the emission from very small, temperature-fluctuating grains \citep{Draine2001, Jones2017}.  Its spectrum varies with the SED of the incident radiation, not with intensity. The spectrum of recurrent fluorescence from carbon clusters also depends on the energy of the exciting photon, but not on the intensity of the excitation radiation \citep{Lacinbala2023PhRvA}.  Therefore, all three possible origins are consistent with the observed lack of a trend in $T_\mathrm{L}$ with distance from IF.

While its spectral shape is not fully constrained in the present study, the high-temperature component at 1--2\,$\mu$m seems secure. It shows a decreasing trend with wavelength, which is different from the free-free emission. Therefore, it is unlikely due to the insufficient subtraction of the free-free emission (Fig.~\ref{fig:average}). Scattered starlight is present over the extended Orion nebula \citep{O'Dell2009}, and the high-temperature component could contain a contribution from scattered light by sub-micrometer-sized dust grains. The diffuse galactic light shows continuum emission for 0.95--1.65\,$\mu$m, which can be attributed to the scattered starlight by dust grains \citep{Arai2015}. Its spectrum can be approximated by $I_\nu \propto \nu^{-1.36}$ for 0.95--1.65\,$\mu$m. The high-temperature component exhibits a slowly increasing trend with wavelength, $I_\nu \propto \nu^p$, where $p > 0$ for 1--1.65\,$\mu$m, since the peak of the fitted blackbody is located around this spectral range. The shorter-wavelength side of the high-temperature component may exhibit a contribution from the scattered light, as seen in the case with the diffuse galactic light; however, there must be another component with a different origin in the longer-wavelength side of the high-temperature component. The present data are not sufficient to further discuss its origin.

\subsection{Absorption features at 3.0 and 4.27\,$\mu$m}\label{subsec:absorption}
Figure~\ref{fig:average} indicates absorption features at around 3.0\,$\mu$m and 4.27\,$\mu$m.  They can be attributed to water ice and CO$_2$ ice.  To confirm the presence of these ice absorption features, we further analyzed these features. See Appendix~\ref{app:ice} for details.  The absorption feature at 3.0\,$\mu$m is well fitted by amorphous water ice absorption (Fig.~\ref{fig:H2O}).  The feature at 4.27\,$\mu$m is weak, and the absorption strength was estimated by integrating the optical depth.  The column densities and abundance ratios were estimated using the assumed band strengths (see Appendix~\ref{app:ice}). They are summarized in Table~\ref{table:ice}, where 1$\sigma$ is given as an uncertainty. The column density of the water ice does not show significant spatial variation over the observed area, except for region 9. 
The detection of the absorption feature at 4.27\,$\mu$m is not significant for regions 1 and 9 and can be considered an upper limit. It is also tentative for regions 2, 5, and 6, whereas for other regions the absorption is detected at more than 4$\sigma$.  The detection of both absorption features supports their assignment to ice grains. A similar water ice absorption feature is also reported in NGC\,7023 \citep{Misselt2025}, which also supports the assignment. The CO$_2$ to H$_2$O abundance ratios range over several percent. They are several to an order of magnitude smaller than the values observed for young stellar objects and quiescent clouds \citep[e.g.,][]{Boogert2015}.  The profile might contain foreground continuum emission that partly fills the absorption features (see below), which underestimates the column density and abundance ratio.

The presence of water ice with $\sim 10^{18}$\,cm$^{-2}$ suggests environments with a high column density \citep[e.g., $A_\mathrm{V} \gtrsim 10$, ][]{Boogert2015}. Since the foreground extinction is only $A_\mathrm{V} = 0.9 - 1.9 $ \citep{Habart2024, Peeters2024}, these ice species must be present in a deeper region of the PDR either in the line of sight or in the background PDR behind the ionized or neutral gas region. Radio observations suggest that the Orion Bar region has at least $N_\mathrm{H} = 1 \times 10^{23}$\,cm$^{-2}$ \citep{Berne2014, Salgado2016}, which can be converted to $A_\mathrm{V} = 35 $ \citep{Habart2024}.  This is sufficient to form ice species. Observations of water vapor also support the presence of abundant water ice in dense clouds of the Orion Molecular Cloud \citep{Melnick2020}. The present detection of ices suggests that a background IR source must exist to produce the absorption features of ice in deeper regions and that the observed NIR continuum emission is superimposed on it.  The foreground continuum may have more emission at 4\,$\mu$m than at 3\,$\mu$m and thus fills the CO$_2$ absorption more, resulting in smaller abundance ratios.  The contribution of the background emission at 3\,$\mu$m can be estimated at at least 40\%, assuming that water ice absorption would saturate at peak wavelengths. The presence of ice absorption suggests that some fraction of the NIR continuum comes from either a deeper region or the background PDR. The spectral shape of the NIR continuum emission may be affected by background emission, but the sharp decline characterized by the blackbody with $T \sim 700$\,K should not significantly change. Further studies of other regions are needed to confirm the spectral shape of the origin of the NIR continuum emission.

\begin{table*}
\caption{Column densities of the ice species.}             
\label{table:ice}      
\centering                          
\begin{tabular}{c c c c c}        
\hline\hline                 
Region \#& Region & H$_2$O ice column density & CO$_2$ ice column density & CO$_2$ ice to H$_2$O ice abundance ratio\\  
 & & ($ \times 10^{17}$\,cm$^{-2}$) &  ($ \times 10^{17}$\,cm$^{-2}$) & (\%) \\
\hline                        
   1 & ionized region & $8.60  \pm 0.16$ & $0.09 \pm 0.07$ & $ 1.0 \pm 0.8$\\
   2 & teal region    & $ 8.42 \pm 0.10$ & $0.15 \pm 0.05$ & $ 1.8 \pm 0.6$  \\
   3 & atomic PDR     & $10.14 \pm 0.07$ & $0.23 \pm 0.05$ & $ 2.3 \pm 0.5$  \\
   4 & DF1            & $10.52 \pm 0.12$ & $0.31 \pm 0.04$ & $ 3.0 \pm 0.4$  \\
   5 & DF1 -- DF2     & $10.46 \pm 0.13$ & $0.23 \pm 0.07$ & $ 2.2 \pm 0.7$ \\ 
   6 & DF2            & $8.91 \pm 0.20$  & $0.29 \pm 0.09$ & $ 3.2 \pm 1.0$ \\
   7 & DF2 -- DF3     & $7.88 \pm 0.19$  & $0.52 \pm 0.06$ & $ 6.6 \pm 0.8$ \\
   8 & DF3            & $11.36\pm 0.17$  & $0.54 \pm 0.06$ & $ 4.8 \pm 0.5$ \\
   9 & beyond DF3     & $14.33\pm 0.28$  & $0.41 \pm 0.17$ & $ 2.9 \pm 1.2$ \\ 
\hline                                   
\end{tabular}
\tablefoot{The uncertainty given is 1$\sigma$.}
\end{table*}

\section{Summary}
\label{sec:summary}
We conducted a detailed analysis of the NIR continuum emission in the Orion Bar using the NIRSpec IFU data from the PDRs4All program \citep{Berne2022, Peeters2024}. Contribution from ionized gas was estimated using the Cloudy code and subtracted. Excess emission remains clearly in the 1--4.5\,$\mu$m range even after subtracting the free-free and free-bound emission contributions from the ionized gas. We obtained the average spectra of nine physically distinct regions. The continuum emission shows good correlations with the 3.3\,$\mu$m AIB intensity at wavelengths longer than 2.7\,$\mu$m. However, the emission at 1.2\,$\mu$m shows poor correlation except for the atomic PDR and the region beyond DF3. We demonstrated that the shape of the NIR continuum can be approximated well by a summation of two blackbodies. The low-temperature component can also be approximated by a modified blackbody, and the difference in the fit is insignificant. The temperature of the low-temperature component ranges from 200--1000\,K with a mean of $\sim 700$\,K for the blackbody fit and $\sim 500$\,K for the modified blackbody fit. The temperature of the high-temperature component ranges higher than 2000\,K and is not well constrained by the present data.  We find that the low-temperature component correlates well with the 3.3\,$\mu$m AIB but that the high-temperature component does not.

Three possible origins of the low-temperature component were discussed: the quasi-continuum by PAH cations \citep{Esposito2024}, nanometer-sized carbonaceous particles \citep[THEMIS model,][]{Jones2017, Elyajouri2024}, and recurrent fluorescence from carbon clusters \citep{Lacinbala2023}. Polycyclic aromatic hydrocarbon cations may contribute to the emission at wavelengths longer than 3\,$\mu$m, but it is not clear whether they emit at shorter wavelengths efficiently. Nanometer-sized carbonaceous particles emit both in the continuum and the 3.3\,$\mu$m band, and it is difficult to fit both with the same temperatures. Emissivity may need to be revised. Recurrent fluorescence from isomers of the carbon clusters, C$_{42}$ and C$_{60}$, produces blackbody-like emission.  The estimated temperature of 747\,K of the recurrent fluorescence for the reflection nebula NGC\,7023 suggests that larger carbon clusters may explain the observed emission of the low-temperature component in the Orion Bar. The carriers of the high-temperature component cannot be unambiguously determined but must fulfill a decreasing trend with wavelength.

The average spectra of emblematic regions explored in the Bar region also suggest the presence of absorption features. 
Absorption features at 3.0 and 4.27\,$\mu$m are securely detected and are attributed to water ice and CO$_2$ ice, respectively. The simultaneous detection of the two ice species provides further support for these assignments. Since the foreground extinction $A_\mathrm{V}$ is less than 2, these absorption features must arise in a deeper region of the PDR in the line of sight, which also contributes to a fraction of the observed NIR continuum. It may contribute to the spectral shape, but the sharp decline characterized by the blackbody with $\sim 700$\,K should not be affected. Further studies of other targets are needed to confirm the spectral shape and further explore the origin of the NIR continuum.

\begin{acknowledgements}
      The authors thank the anonymous reviewer for many useful comments.  They also thank Baria Khan for providing the region designation file and Meriem Elyajouri for the data of the THEMIS model.
      This work is based on observations made with the NASA/ESA/CSA James Webb Space Telescope. The data were obtained from the Mikulski Archive for Space Telescopes at the Space Telescope Science Institute, which is operated by the Association of Universities for Research in Astronomy, Inc., under NASA contract NAS 5-03127 for JWST. These observations are associated with programme \#1288 (DOI: 10.17909/pg4c-1737).
      
      Support for programme \#1288 was provided by NASA through a grant from the Space Telescope Science Institute, which is operated by the Association of Universities for Research in Astronomy, Inc., under NASA contract NAS 5-03127.

      TO acknowledges the support by the Japan Society for the Promotion of Science (JSPS) KAKENHI Grant Number JP24K07087. 

      CB is grateful for an appointment at NASA Ames Research Center through the San Jos\'e State University Research Foundation (80NSSC22M0107) and acknowledges support from the Internal Scientist Funding Model (ISFM) Laboratory Astrophysics Directed Work Package Rd3 at Ames.

      YO acknowledges the Collaborative Research Center 1601 (sub-project A6 and C3) funded by the Deutsche Forschungsgemeinschaft (DFG, German Research Foundation) - Project-ID 500700252 - SFB 1601. 

      EP acknowledges support from the University of Western Ontario, the Canadian Space Agency (CSA, 22JWGO1-16), and the Natural Sciences and Engineering Research Council of Canada.

      JRG thanks the Spanish MCINN for funding support under grant PID2023-146667NB-I00.

      This project has received funding from the European Research Council (ERC) under the European Union's Horizon Europe research and innovation programme ERC-AdG-2022 (SUL4LIFE GA No. 101096293).  AF also thanks project PID2022-137980NB-I00 funded by the Spanish Ministry of Science and Innovation/State Agency of Research MCIN/AEI/10.13039/501100011033 and by "ERDF A way of making Europe".
\end{acknowledgements}


\bibliographystyle{aa}
\bibliography{NIRcont}

\begin{appendix}
\onecolumn
\section{Cloudy simulations} \label{app:cloudy}
\begin{figure*}[tbh]
   \centering
   \includegraphics[width=\hsize]{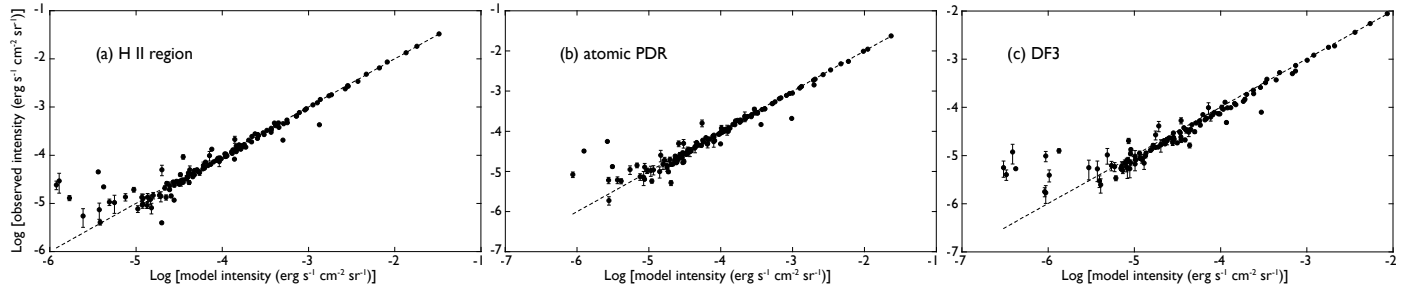}
      \caption{Observed line intensities are plotted against the model estimates.  120 lines detected in the NIRSpec and MIRI MRS spectra are plotted.  (a) ionized region, (b) atomic PDR, and (c) DF3. The six strongest lines used for the fit are \ion{H}{i}\xspace Pa\,$\alpha$, \ion{He}{i}\xspace 1.0834\,$\mu$m, \ion{H}{i}\xspace Pa\,$\beta$, \ion{H}{i}\xspace Br\,$\alpha$, \ion{H}{i}\xspace Pa$\,\gamma$, and \ion{H}{i}\xspace Br\,$\beta$ from the right top.  The dashed lines show the best linear-fit lines using all the lines plotted.
              }
         \label{fig:line-fit}
   \end{figure*}

We adopted the same model setup as in \citet{Shaw2009} and \citet{Pellegrini2009}.  We assumed an open geometry and the central source was given by a Kurucz stellar atmosphere model with an effective temperature of 39,600\,K \citep{Shaw2009, Pellegrini2009}.  The incident radiation from a hot bremsstrahlung of a temperature of 10$^6$\,K, as indicated by X-ray observations, was also assumed \citep{Feigelson2005, Pellegrini2009}.  For the fit to the template spectra, we used the six strongest lines from the ionized gas in the Orion Bar region, \ion{H}{i}\xspace Paschen (Pa) $\alpha$, $\beta$, and $\gamma$ at 1.8756, 1.2822, and 1.0941\,$\mu$m, and Bracket (Br) $\alpha$ and $\beta$ at 4.0525, and 2.6260\,$\mu$m, respectively, and \ion{He}{i}\xspace ($^3$S--$^3$P$_0$) at 1.0834\,$\mu$m to search for the model parameters that best matched the observed intensities of these lines after applying the foreground extinction described in Sect.~\ref{sec:obs} on the intensities predicted by Cloudy.  
We set the gas number density $n$ and the flux of hydrogen-ionizing photons striking the face of the cloud $\phi$ as free parameters and found their best fit values.  We first roughly searched in the range $n = 10^2 - 2.5 \times 10^5$\,cm$^{-3}$ and $\phi = 4 \times 10^{11} - 2 \times 10^{14}$\,cm$^{-2}$\,s$^{-1}$ that centered on the values of the IF (3200\,cm$^{-3}$ and $6.45 \times 10^{13}$\,cm$^{-2}$\,s$^{-1}$) as estimated previously by \citet{Pellegrini2009} and \citet{Peeters2024}, with a logarithmic step size of 0.5.  We then narrowed the best parameter range down to 0.01, on a logarithmic scale, by minimizing the reduced chi-square, $\chi^2 = \Sigma((I_\mathrm{obs} - I_\mathrm{Cloudy})/\sigma)^2/N$, where $I_\mathrm{obs}$ and $I_\mathrm{Cloudy}$ are the observed and Cloudy output line intensities, respectively, and $\sigma$ is the uncertainty in the observed intensity and $N$ the number of lines in the fit.  The best-fit parameters obtained were estimated as ($n$, $\phi$) = (10$^{3.96}$, 10$^{13.0}$), (10$^{3.68}$, 10$^{12.89}$), and (10$^{3.59}$, 10$^{12.34}$) for the \ion{H}{ii}\xspace, atomic PDR, and DF3 template spectra, respectively, where $n$ is in units of cm$^{-3}$ and $\phi$ is in units of cm$^{-2}$\,s$^{-1}$. Figure~\ref{fig:line-fit} shows the plots of the observed line intensities versus model estimates, including weaker lines of \ion{H}{i}\xspace and \ion{He}{i}\xspace in the NIRSpec and MIRI/MRS spectra taken from \citet{Peeters2024} and \citet{Vandeputte2024}.  The model estimates agree well with the observed line intensities for the three template spectra. The observed-to-model intensity ratios of all the lines plotted are $1.0039\pm 0.0001$, $0.9981\pm 0.0001$, and $1.0146 \pm 0.00005$, for the \ion{H}{ii}\xspace, atomic PDR, and DF3, respectively.  The other two template data (DF1 and DF2) also show a similarly good agreement with the model estimates.  The estimated values for the \ion{H}{ii}\xspace region are consistent with those derived for the ionized region from previous studies \citep{Walmsley2000, Pellegrini2009, Weilbacher2015, Peeters2024}, suggesting that the present fit is reliable and that the ionized gas toward the atomic and molecular PDRs exists in the vicinity of the Bar region.

We extended the fitting to the entire NIRSpec region.  
Since \ion{H}{i}\xspace Br\,$\alpha$ falls at the edge of the spectral gap, it is observed for about 65\% of the area.  To make a consistent analysis, we employed the other five lines to perform Cloudy simulation by minimizing the reduced chi-squared value, excluding Br\,$\alpha$. After searching the best fit parameters for the range $n = 10^2 - 2.5 \times 10^5$\,cm$^{-3}$ and $\phi = 4 \times 10^{11} - 2 \times 10^{14}$\,cm$^{-2}$\,s$^{-1}$ as described above, we found that the data were all best fitted within $n = 10^{2.5} - 10^{4.4}$\,cm$^{-3}$ and $\phi$ = $10^{12.1} - 10^{13.3}$\,cm$^{-2}$\,s$^{-1}$.  We searched for the best values in this range with  logarithmic steps of 0.01.  A typical value of $\chi^2$ in the best fit is about 3, suggesting that the fit is not perfect, but reasonable.  The deviations of 0.02 on a logarithmic scale in both parameters increases $\chi$ by more than unity, which makes a change in the free-free and free-bound intensity by 1.6, 2.0, 2.4, 2.0, and 2.7 MJy\,sr$^{-1}$ on average, for the continuum at 1.2, 2.1, 2.7, 3.8, and 4.3\,$\mu$m, respectively.  We regard these values as typical uncertainties in the present estimate. 

Br\,$\alpha$ is the fourth strongest line in the Orion Bar region, and we investigated the effect of its exclusion by comparing the fit results with and without it for the regions where it is observed.  Table~\ref{table:compBra} summarizes the results.  The difference in the surface brightness increases slightly with wavelength, and is the largest at 4.3\,$\mu$m with $0.5 \pm 0.8$\,MJy\,sr$^{-1}$.  These values are a factor of 4--6 smaller than the uncertainty in the line fit, as described above.  The difference relative to the results without Br\,$\alpha$ decreases with wavelength as the contribution of free-free emission also decreases and is the largest at 1.2\,$\mu$m with $2.4 \pm 10.8$\%.  The pixel-level difference shows large scatter, but the general trend is not changed by the exclusion of Br\,$\alpha$.  The exclusion affects the Spearman correlation coefficients with the 3.3\,$\mu$m AIB only slightly.

\begin{table*}
\caption{Comparison of the analysis with and without \ion{H}{i}\xspace Br\,$\alpha$ }         
\label{table:compBra}      
\centering                          
\begin{tabular}{c c c c c c c}        
\hline\hline                 
& & Cont1.2 & Cont2.1 & Cont2.7 & Cont3.8 & Cont 4.3  \\    
\hline                     
\multicolumn{2}{c}{difference (MJy\,sr$^{-1}$)} & $0.31 \pm 0.42 $ & $0.38 \pm 0.50 $ & $0.38 \pm 0.52 $ & $0.45 \pm 0.63 $ & $0.49 \pm 0.81 $\\
\multicolumn{2}{c}{difference (\%)} & $2.4 \pm 10.8 $ & $1.5 \pm 2.0 $ & $1.1 \pm 3.3 $ & $0.3 \pm 0.6 $ & $0.3 \pm 0.7 $\\
\hline
Region \# &  Br\,$\alpha$ & \multicolumn{5}{c}{Spearman correlation coefficient with the 3.3\,$\mu$m AIB intensity}\\
\hline
   1 & with    & 0.35 & 0.26 & 0.57& 0.95 & 0.74   \\ 
        \vspace{3pt}
     & without & 0.33 & 0.23 & 0.57 & 0.95 & 0.71  \\  
   2 & with    & 0.21 & 0.75 & 0.71 & 0.80 & 0.78   \\
    \vspace{3pt}
     & without & 0.17 & 0.74 & 0.71 & 0.80 & 0.78  \\
   3 & with    & 0.45 & 0.79 & 0.65 & 0.90 & 0.91  \\
    \vspace{3pt}
     & without & 0.44 & 0.78 & 0.64 & 0.90 & 0.91  \\
   4 & with    & 0.02 & 0.04 & 0.42 & 0.74 & 0.76  \\
    \vspace{3pt}
     & without & 0.03 & 0.03 & 0.43 & 0.78 & 0.80  \\
   5 & with    & 0.35 & 0.07 & 0.87 & 0.96 & 0.97  \\
    \vspace{3pt}
     & without & 0.32 & 0.02 & 0.85 & 0.96 & 0.96  \\
   6 & with    & 0.26 & 0.50 & 0.72 & 0.95 & 0.95  \\
    \vspace{3pt}
     & without & 0.26 & 0.51 & 0.72 & 0.95 & 0.94  \\
   7 & with    &-0.17 & 0.33 &-0.32 & 0.61 & 0.68  \\
    \vspace{3pt}
     & without &-0.18 & 0.32 &-0.32 & 0.60 & 0.68  \\
   8 & with    &-0.03 &-0.27 & 0.43 & 0.95 & 0.93  \\
    \vspace{3pt}
     & without &-0.02 &-0.27 & 0.42 & 0.95 & 0.93  \\
   9 & with    & 0.68 & 0.41 & 0.65 & 0.92 & 0.62  \\
     & without & 0.67 & 0.35 & 0.63 & 0.90 & 0.60  \\
   \hline
\end{tabular}
\end{table*}

\section{Removal of emission lines}\label{app:removal}

\begin{figure*}[hb]
   \centering
   \includegraphics[width=\hsize]{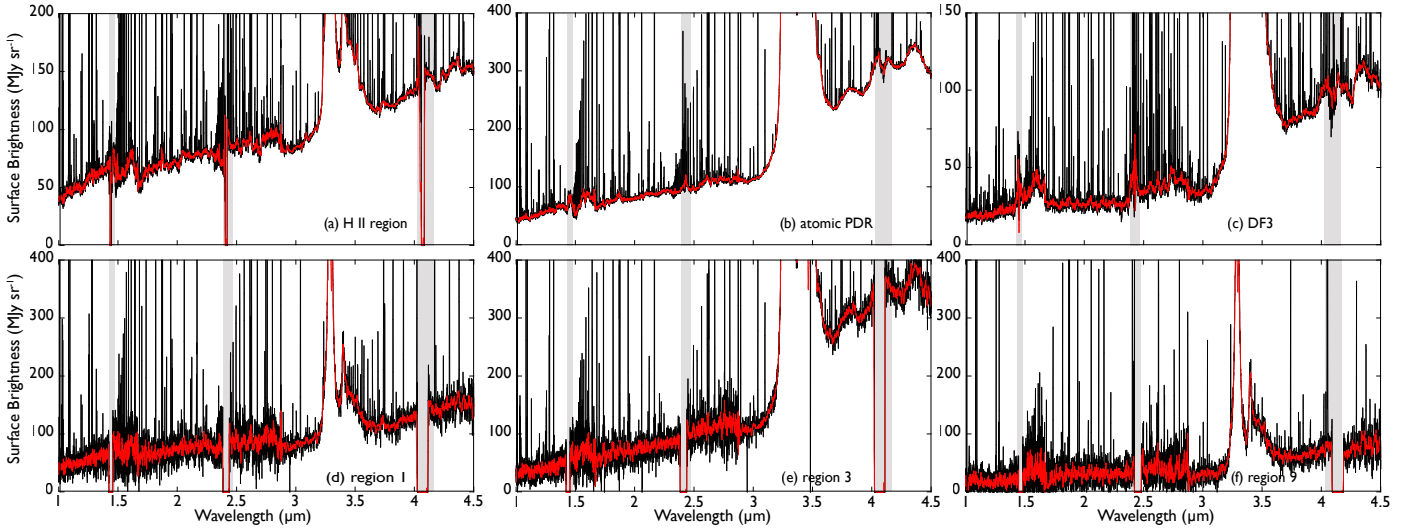}
      \caption{Line removal process applied to the template spectra (top row) and spectra of single spaxels (bottom row).  The black lines show the original observed spectra, while the red lines indicate the continuum after applying the 3-$\sigma$ clipping using a 32-point averaging.  The top row shows (a) \ion{H}{ii}\xspace region, (b) atomic PDR, and (c) DF3 templates.  The bottom row presents spectra of single spaxels, randomly chosen from (d) region 1, (e) region 3, and (f) region 9. The shaded regions indicate the spectral gaps in the observations, where the data are not reliable.} 
         \label{fig:lineremoval}
   \end{figure*}

The NIR spectra in the Orion Bar region show a number of emission lines \citep{Peeters2024}, which need to be removed to present the continuum component clearly. We calculated an average over 32 data points in the spectral direction and removed the data points that deviate more than 3$\sigma$ from the average.  The top row of Figure~\ref{fig:lineremoval} shows this sigma-clipping applied to the three template spectra. The black lines show the original observed spectra and the red lines indicate the continuum emission after applying the 3-$\sigma$ clipping process.  We also checked the efficiency of this process at the single spaxel level.  The bottom row of Figure~\ref{fig:lineremoval} presents the spectra of single spaxels randomly selected from regions 1, 3, and 9.  We inspected individual spectra and verified that this line removal process works well also at the spaxel level. Spaxels severely affected by the noise 1/f were excluded by the 5$\sigma$ clipping process (see Sect.~\ref{subsec:spectrum}).

\section{Two-component fit results}\label{app:2body}

\begin{figure*}[hb]
   \centering
   \includegraphics[width=\hsize]{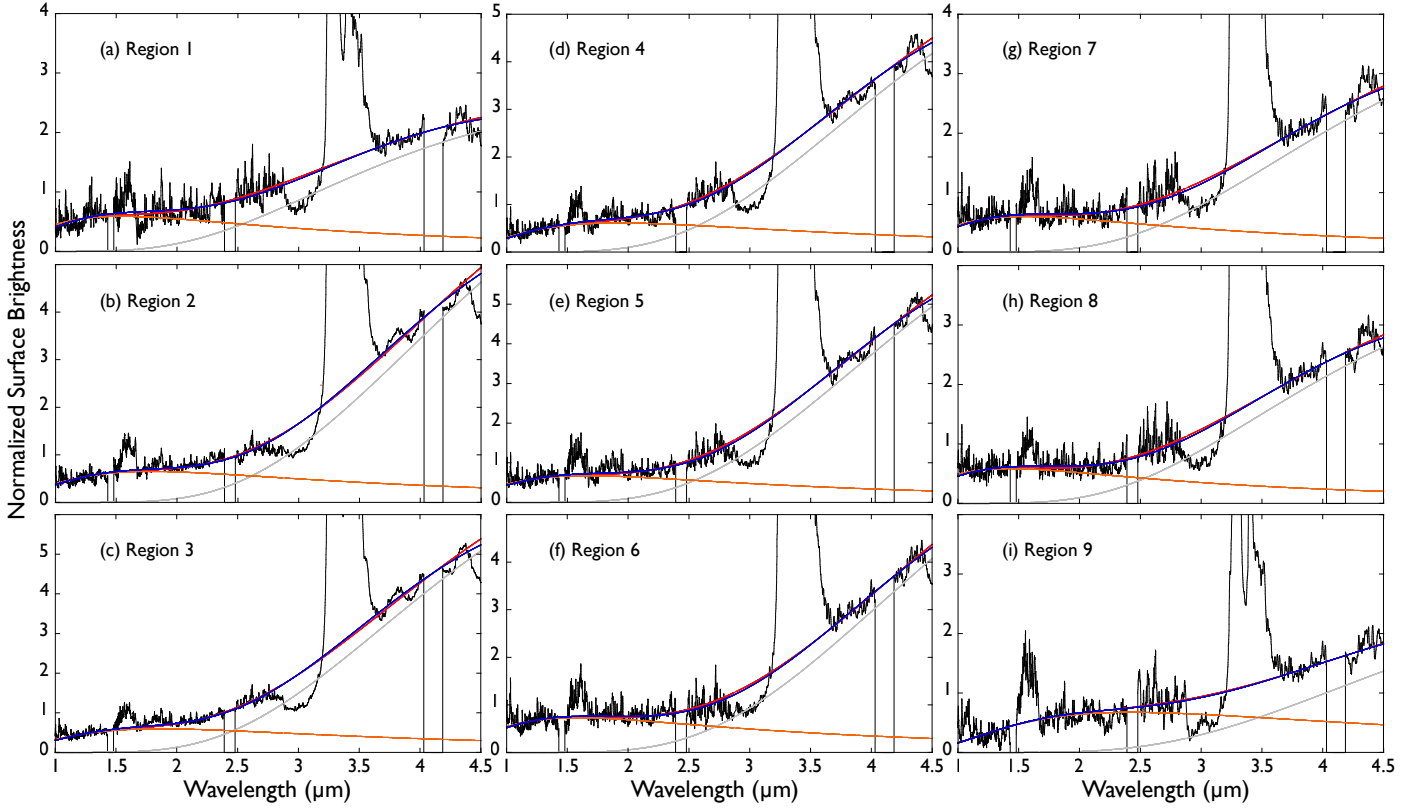}
      \caption{Two-blackbody and modified black-body fit for the average spectra of the nine regions. The black line shows the average spectra.  The solid red and blue lines indicate a fit with two blackbodies (Equation~(\ref{eq1})) and a fit with a modified blackbody and blackbody (Equation~(\ref{eq4})), respectively. The gray line and light brown line indicate the low- and high-temperature components for the two-blackbody fit.  Note that the red and blue lines mostly overlap with each other.  (a), (b), ... (h) show regions 1, 2, ..., and 9, respectively.}  
         \label{fig:BBsp}
   \end{figure*}

\begin{figure*}[tbh]
   \centering
   \includegraphics[width=0.55\hsize]{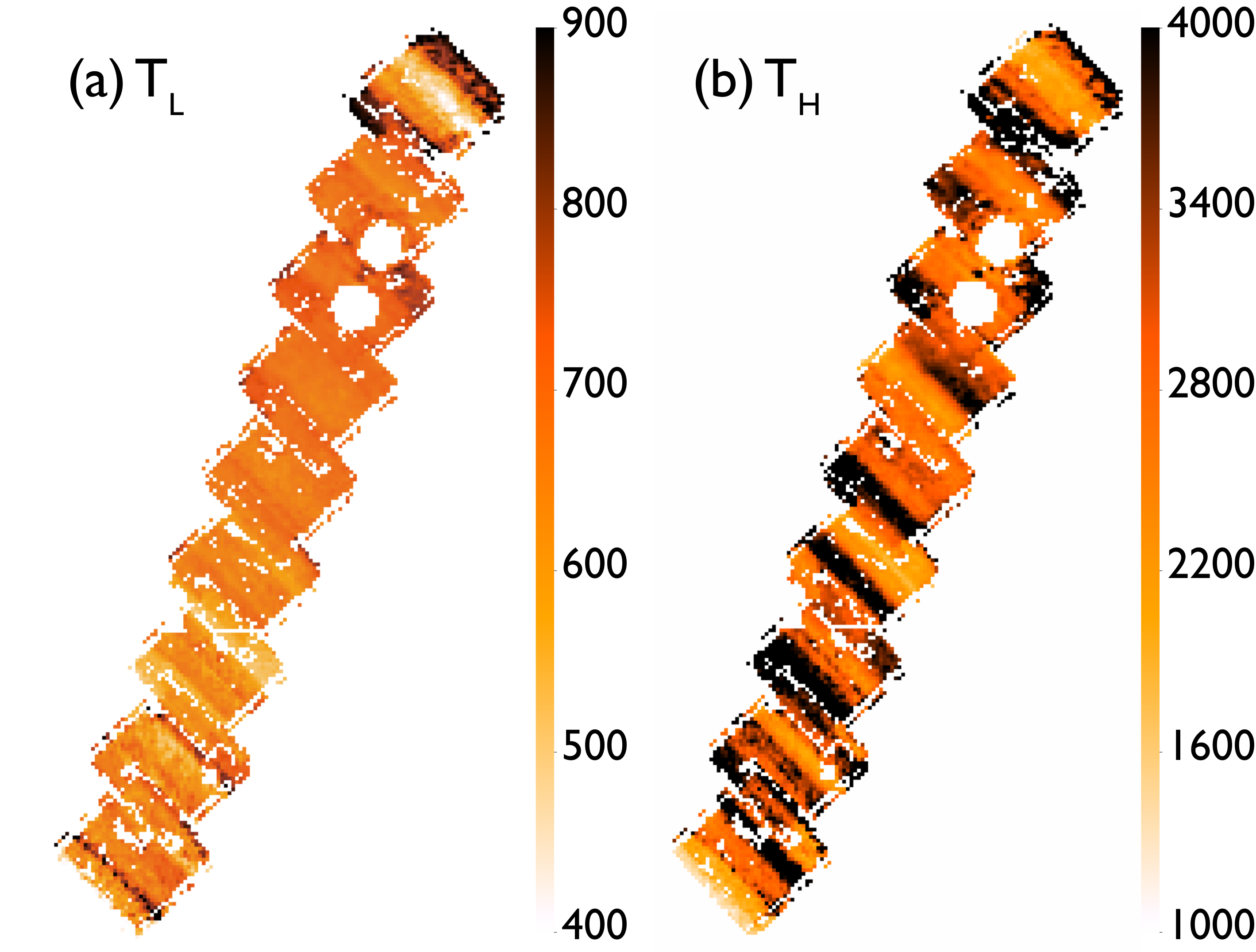}
      \caption{Maps of temperatures of the low- and high-temperature components, $T_\mathrm{L}$ (a) and $T_\mathrm{T}$ (b) of the two-blackbody fit.}
         \label{fig:Tdist}
   \end{figure*}

Figure~\ref{fig:BBsp} shows the results of the two-blackbody fit (Equation~(\ref{eq1})) by the solid red lines and the modified blackbody and blackbody fit (Equation~(\ref{eq4})) by the solid blue lines to the average spectra of the nine regions together with each temperature component.  Figure~\ref{fig:Tdist} shows the maps of the two fitted temperatures of for the entire NIRSpec region.

\begin{figure*}[tbh]
   \centering
   \includegraphics[width=\hsize]{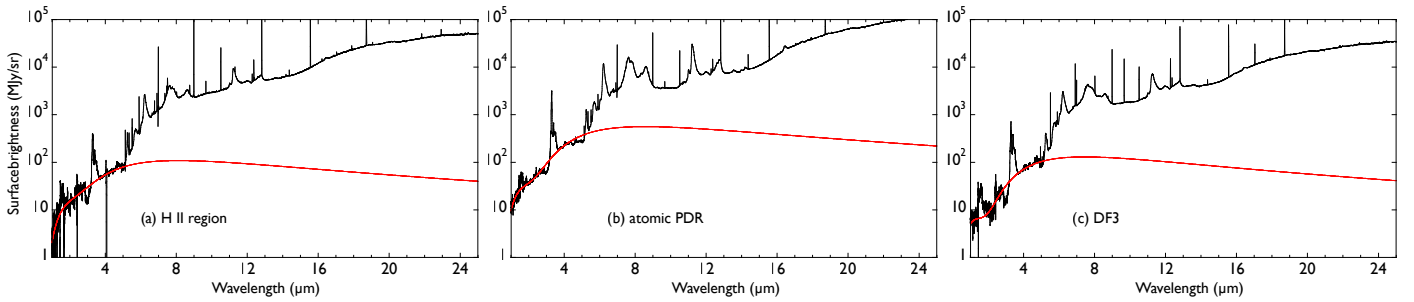}
      \caption{Observed NIRSpec and MIRI/MRS spectra of three templates after subtracting the contribution from the ionized gas and correcting the extinction (black lines) and the two blackbody fit spectra (red lines).  For the NIRSpec spectra, emission lines are removed as described in Appendix~\ref{app:removal}, while they are not in the MRS spectra. }  
         \label{fig:N+M}
   \end{figure*}

The low-temperature component may exceed the observed MIR spectra taken with MIRI/MRS \citep{Chown2024}.  Since the NIRSpec and MRS have a different pixel scale, we used the template spectra, which were extracted from the same area \citep{Chown2024}, to compare the two-blackbody fit with the MRS spectra.  Figure~\ref{fig:N+M} plots the NIRSpec and MRS spectra of the three templates together with the two-blackbody fits.  The contribution from the ionized gas was subtracted and the extinction was corrected for both the NIRSpec and MRS spectra.  The fitted temperatures ($T_\mathrm{H}$, $T_\mathrm{L})$ are (1803\,K, 605\,K), (2197\,K, 590\,K), and (3417\,K, 682\,K), for \ion{H}{ii}\xspace region, atomic PDR, and DF3, respectively.  As shown in Figure~\ref{fig:N+M}, the MRS spectra show a rapid increase from 5\,$\mu$m and the two-blackbody fit stays well below the MRS spectra.  Therefore, the two-blackbody fit does not conflict with the MRS spectra.  The modified blackbody fit has a higher temperature and decreases more steeply at longer wavelengths due to the factor $\lambda^{-2}$, and does not conflict with the MRS spectra either.

%

\section{Absorption features at 3.0 and 4.27\,$\mu$m}\label{app:ice}
\begin{figure*}
   \centering
   \includegraphics[width=\hsize]{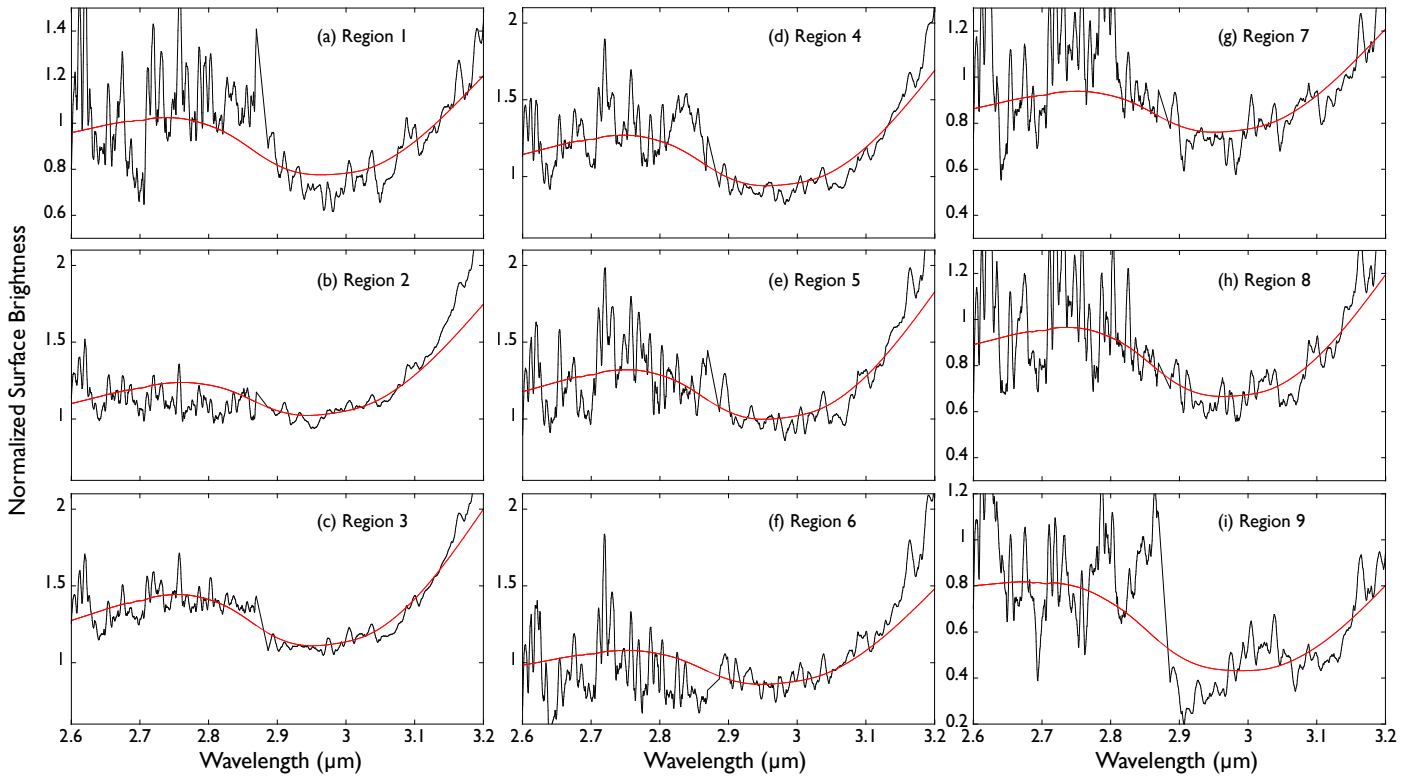}
      \caption{Enlarged average spectra in the range 2.6--3.2\,$\mu$m in black and fit with amorphous water ice at 15\,K in red \citep{Mastrapa2009}.  The observed spectral are smoothed by the Savitzky-Golay filter.  The baseline continuum is assumed to be given by the two-blackbody fit.  The assignment of the region is the same as in Figure~\ref{fig:average}}.  
         \label{fig:H2O}
   \end{figure*}

Figure~\ref{fig:H2O} shows the 3\,$\mu$m region of the average spectra and the fit with the water ice absorption.  The observed spectra were smoothed by the Savitzky-Golay filter (see Sect.~\ref{sec:f-f}).  For the water ice absorption, we employed the optical constant of amorphous water ice at 15\,K provided by \citet{Mastrapa2009} and assumed the weighted continuous distribution of ellipsoids for the shape \citep{Ossenkopf1992}.  The continuum level was assumed to be given by the two-blackbody fit.  For all the regions, the fit is good except for the longer wavelength side, where the 3.3\,$\mu$m AIB emission starts dominating and the baseline continuum is not well approximated by the two-blackbody fit.
The fit results show that the feature can be attributed to the water ice absorption.
The column density of water ice was estimated, assuming that the band strength is $1.9 \times 10^{-16}$\,cm\,molecule$^{-1}$ \citep{Mastrapa2009}.

\begin{figure*}
   \centering
   \includegraphics[width=\hsize]{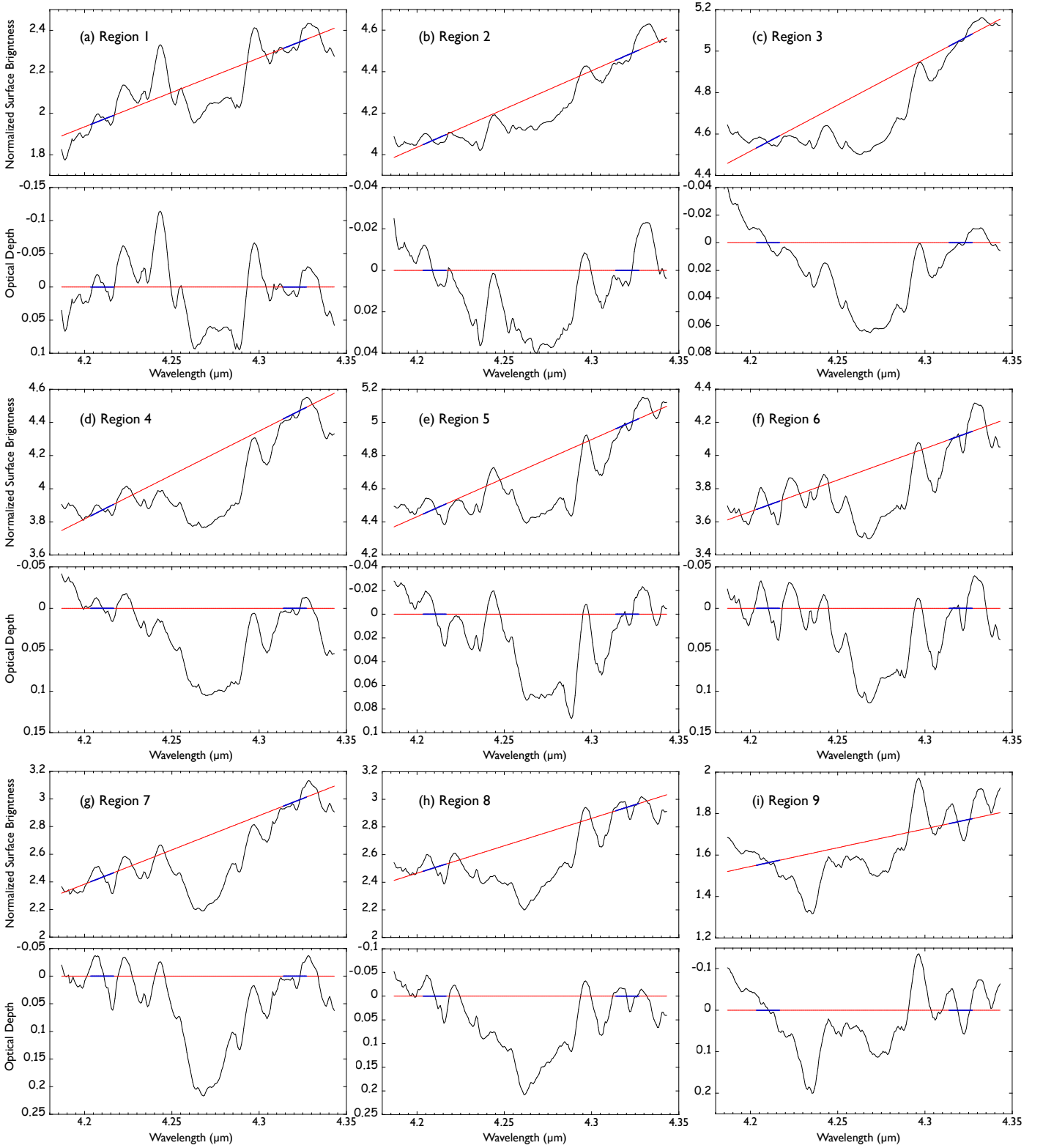}
      \caption{Enlarged average spectra in the range 4.2--4.34\,$\mu$m (black) and the assumed linear continuum (red) are plotted in the upper panels for 9 regions.  The spectra are smoothed using a Savitzky-Golay filter.  The fiducial continuum regions are indicated by the thick blue lines. The lower panels show the established optical depth from the assumed continuum, where the optical depth = 0 (assumed continuum) and the fiducial continuum regions are indicated by the red and blue lines, respectively.  The region assignment is as in Figure~\ref{fig:average}.}
         \label{fig:CO2}
   \end{figure*}

The upper panels of Figure~\ref{fig:CO2} show the 4.18--4.34\,$\mu$m average spectra for each of the 9 regions.  The spectra were smoothed using a Savitzky-Golay filter (see Sect.~\ref{sec:f-f}).  Since this spectral range is noisy and the absorption feature at 4.27\,$\mu$m is narrower than the water ice feature, the two-blackbody fit does not provide an appropriate reference continuum.  Therefore, the continuum was locally assumed as a straight line between 4.21 and 4.32\,$\mu$m, as indicated by the red line. The continuum regions were chosen to sufficiently cover the CO$_2$ absorption band observed in dense clouds \citep{Gibb2004, Boogert2015}.  The continuum fiducial points were calculated as an average over 21 data points on each side, which are indicated by the thick blue lines. The lower panels indicate the optical depth calculated using the assumed continuum.  The absorption strength was computed by integrating the optical depth bracketed by the blue regions.  As Figure~\ref{fig:CO2} shows, there are "fringe-like" patters remaining in the spectra, and we found that the selection of the continuum region affects the results, when the absorption is weak.  To estimate the systematic effect of the fringe pattern on the derived absorption strength, we changed the assumed continuum range by $\pm 0.01$\,$\mu$m, which roughly corresponds to the period of the fringe pattern, on both sides to estimate the uncertainty on the integrated strength of the feature.  We estimated the column density of CO$_2$ ice, assuming that the band strength is given by $7.6 \times 10^{-17}$\,cm\,molecule$^{-1}$ \citep{Gerakines1995}.  The results are summarized in Table~\ref{table:ice}, which includes one $\sigma$ uncertainty derived in the assumed continuum.

   \end{appendix}

\label{LastPage}
\end{document}